\documentclass[a4paper,conference]{IEEEtran}
 \usepackage{graphicx} 
 \usepackage[cmex10]{amsmath} 
 \usepackage{amssymb,verbatim}

\begin{document}
%
\title{%
 Quantum estimation theory with quantum noise control\\
}

\author{
\IEEEauthorblockN{ Osamu Hirota$^{1,2}$, Masaki Sohma$^{1}$ \\}
\IEEEauthorblockA{
1. Quantum ICT Research Institute, Tamagawa University\\
6-1-1, Tamagawa-gakuen, Machida, Tokyo 194-8610, Japan\\
2. Optical Quantum Information Technology Laboratory Company\\
Totsuka-ward, Yokohama, Kanagawa Pref. Japan\\
{\footnotesize\tt E-mail: hirota@lab.tamagawa.ac.jp, sohma@eng.tamagawa.ac.jp} \vspace*{-2.64ex}}
}

\maketitle

\begin{abstract}
Quantum estimation theory serves as one of  fundamental backbones of quantum Shannon information theory and was mathematically systematized 
by pioneers of quantum information science. Although its applied research saw limited progress for a long period, the 21st century has witnessed 
active discussions and numerous fruitful proposals regarding its applications. Notably, the contributions of Monras and Paris represent one of 
the most remarkable achievements addressing the realization problem of estimation bounds. This paper aims to demonstrate a practical direction 
for extending their original ideas. Next-generation quantum information mechanisms urgently demand guarantees of quantum advantage unattainable 
by classical theory, alongside strict real-time processing without delays. By leveraging the concept of generalized heterodyne detection proposed 
nearly half a century ago, we demonstrate that a scheme based on their results can effectively meet the stringent requirements of 
near-future information technologies.

\end{abstract}

%
\IEEEpeerreviewmaketitle

\section{\textbf{Introduction}}
The communication theory that formed the foundation of information science was designated as "statistical communication theory" by Wiener and 
his colleagues [1]. Their theory is grounded in axiomatic probability theory. It was subsequently advanced by Middleton [2], van Trees [3], 
and others into a design theory for real-time decision and estimation devices, incorporating finite-time or single-shot signal processing for 
rapidly varying signals and noise. Furthermore, Shannon-type information theory, which jointly designs transmitting signals and receiving systems, 
was developed [4]. 

This further refined the probabilistic design theory and led to the blossoming of today’s digital information age. Inspired by these success stories, 
efforts to extend the theory to quantum systems began in the 1960s, and today this has been systematized as quantum Shannon information theory [5, 6]. 
Currently, the primary focus is on developing information technologies based on this theory that can be utilized in the real world.

Quantum mechanics is axiomatized on the basis of quantum probability theory, which elevates axiomatic probability to a probability theory grounded in 
quantum logic, thereby establishing a new dynamical framework. For this reason, the pioneers of quantum communication theory modeled their work after
 communication theory based on axiomatic probability, and constructed a communication theory grounded in quantum probability [7–11]. 

The fundamental concept of quantum communication theory is to formulate signal detection theory, parameter estimation theory, 
and filtering theory using the mathematics of quantum probability. In doing so, they associated Wald’s notion of decision functions [12] 
with a mathematical structure in Hilbert space theory known as decision operators. These decision operators are mathematically generalized resolutions of 
the identity, and in integration theory came to be known as POVM (Positive operator valued measure).

 In quantum communication theory, the central problem is the optimization of decision operators described as POVM. 
 The decision theory defined on quantum systems is  formalized in a manner rooted in communication theory based on axiomatic probability, 
 characterized by the fact that the probability distribution  available to the observer can be controlled through the POVM. 
 Thus, it is important to note that the primary purpose of this theory is to indicate the optimal design of instantaneous decision mechanisms for uncertain 
 phenomena, under constraints such as signal energy and signal processing time, within signal processing systems
  that include the decision process. 
  
However, despite its elegant mathematical structure, the theory left unresolved the question of what physical meaning the optimal POVM  as the fundamental subject of the above theory carries 
and how it can be realized. This issue is known as the POVM realization problem.

Meanwhile, attention turned to the fact that the probability distribution available to the observer also depends on the signal quantum state, 
and research on controlling signal quantum states began in 1974. 
The first challenge in this direction was Yuen’s theory of the two photon coherent state (TCS) [13–15], later referred to as the squeezed state. 
TCS enables control of balance of the uncertainty of non-commuting observables, making it possible to freely design the quantum limits of 
information processing. 
In contrast, in 1977 Hirota proposed introducing degrees of freedom for controlling the uncertainty balance relation into the POVM of 
the measurement process [16] (See appendix A, B, and C). 

In the field of applied research on quantum communication theory, design theories for digital quantum communication systems were actively discussed 
during the 1990s [17–30].
On the other hand, in recent years, based on the early works of Helstrom [7], Holevo [31], Personick [32], and Yuen [33], 
the application of quantum estimation theory dealing with analog information has come to be discussed in various fields [34–43]. In particular, 
numerous examples of the application of the quantum Cramér-Rao inequality have been published. All of these papers have made significant 
contributions to the challenges of the optimal design of analog information processing.

Notably, research such as the phase estimation of optical signals and the simultaneous estimation of non-commuting optical signals can be regarded 
as achievements directly applicable to the development of real-world physical experimental apparatuses. Among them, the studies by Monras [34] 
and Paris [35] provided a major step forward in demonstrating the utility of quantum estimation theory.

We have long been dedicated to the advancement of this field. Building upon the contributions of Monras et al., this paper aims to clearly demonstrate that 
the utility of quantum estimation theory can be further expanded by incorporating the quantum noise control mechanism described in 
Reference [16] into the theory.

Sections 2 and 3 provide a survey of the development of the theory to date. Section 4 outlines the framework of the theory of quantum measurement separation, which is the main subject of this paper. Sections 6 and 7 demonstrate applications to the phase estimation problem.
These serve as examples illustrating the essential need to integrate quantum design principles with classical system design principles, as shown in Figure 1.

\begin{figure}
\centering{\includegraphics[width=8cm]{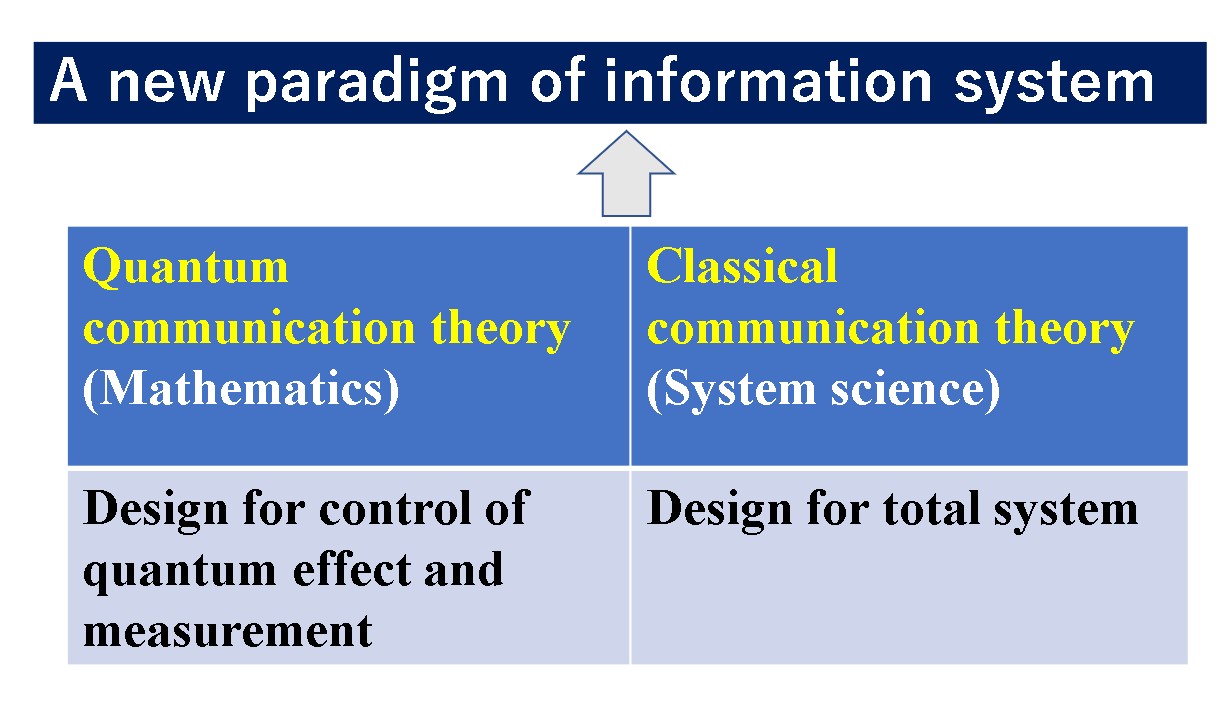}}
\caption{\textbf {Design concept of a new paradigm of information system which exhibits quantum advantage due to quantum effects.}}
\end{figure}

\section{\textbf{ Fundamental Structure of the Quantum Measurement Process}}
\subsection {\textbf {Foundation for quantum communication theory}}

\subsubsection {\textbf {State and measurement}}
Quantum communication theory was developed between the 1960s and 1970s with the primary objective of elucidating the fundamental physical limits 
of optical communication performance. 
In quantum systems, the quantum state or density operator form and measurement of signals are of paramount importance. 

The set of density operator $\rho$ is a convex set, its extreme points being the one-dimensional projection. 
The corresponding states are called pure. Thus the pure state is the most essential for the quantum uncertainty.
Any measurement with values in the real number space $R_M$ is described 
by an affine map of the set of the states (density operator) into the set of probability distributions on $R_M$.
It describes the characteristics of quantum noise.

On the other hand, the measurement process remains one of the most challenging problems 
in quantum mechanics, originating from Von Neumann’s projection postulate. \\

$\textbf {Definition 1}:$
Let us assume that the quantum system has the self adjoint operator as physical observable  
${\bf T}$ and its quantum state $\rho $. 
The Born rule in the standard quantum measurement of the observable is described as follow:
\begin{equation}
{\bf T}|x >=x|x >, P(x) = Tr \rho | x >< x |
\end{equation}
where $| x >< x |$ is called projection valued measure (PVM) in integral theory.\\

The most generalized mathematical description of this process is the theory of generalized resolutions of identity 
or Positive Operator-Valued Measures (POVM). 
Finally the generalized quantum measurement process is described by the following form.\\

 $\textbf {Definition 2}:$
Let us consider a generalized resolution of identity $\{{\bf X}(r) ; r \in R_M \}$, i.e., the collections of Hermitian operators, 
satisfying $\sum_{r\in R_M}{\bf X}(r) =I, {\bf X}(r) \ge 0$.
$Tr \rho {\bf X}(r)$ establishes the one-to-one correspondence between affine maps of the set of density operators into
 the set of probability distributions on $R_M$ and the resolution of identities.
In integration theory, ${\bf X}(r)$ is also called the positive operator valued measure (POVM) based on ${\bf X}(\Delta) = \int_{\Delta} d{\bf X}(r)$.\\
 
 This formulation makes it possible to formulate the Born effect in quantum measurements 
 without going through concrete physical quantities (observable). 
 Davies-Lewis [44],  Benioff[45] and others have been discussing the mathematical generalization 
of the projection process to describe the diversity of measurement processes. 
 \\

\subsubsection {\textbf {Quantum communication theory}}
In the quantum communication theory, the POVM is used as decision operator : $\{\Pi_m\}$ by following operational correspondence.

Let us assume the set $\rho_m, m=\{1,2,3,\dots, M\}$ of quantum states.
 The decision result through the quantum measurement is described by a compact set of the decision operator: 
$\Pi_l$, $l=1,2,3, \dots ,M$ such that 
\begin{subequations}
\begin{equation}
P(l|m)=Tr \rho_m \Pi_l, \quad m,l=1,2,3,\dots, M 
\end{equation}
\begin{equation}
\sum^{M}_{l=1} \Pi_l =I, \quad \Pi_l \ge 0 \quad \forall l
\end{equation}
\end{subequations}
where $I$ is the identity operator.
The decision operator does not just represent the probability of a quantum measurement process, but the probability of 
error or detection by the receiver's decision.
 In other words, it must be understood that it involves a decision by observer [5-11].\\

Thus, the goal of quantum communication theory is to design a quantum-optimal receiving system for quantum noise in the quantum measurement process. 
This is completed by deriving the optimal  decision operator and physically realizing it. 
The subsequent total system design then becomes the role of conventional communication theory.
Specifically, we introduce the following three models based on the relationship between measurement and decision theory (Fig.2).

These quantum measurement processes as communication channels encompasses Quantum Detection Theory
 for discrete signal systems and Quantum Estimation Theory for continuous signal systems. 
 In each, there exist methodologies for individual measurement  and collective measurement. 
 While this paper primarily focuses on continuous systems, we first briefly explain discrete systems to provide a basis for comparison.\\
\textbf{(a) Discrete system (Digital)}\\
Quantum decision theory for discrete signals is known as Quantum Detection Theory. Here, a significant difference emerges between Model B and Model C. 
The POVM in Model C is defined as a quantum version of Wald’s decision function, possessing the dual functions of measurement and decision. 
Its optimal solution under the quantum Bayes criterion yields a "quantum gain." Typical examples include the Helstrom bound for binary signals [7] and 
the Yuen optimal solution for coding signals [9].\\
\textbf{(b) Continuous system (Analog)}\\
When signals are continuous, the problem becomes one of Quantum Estimation. Here, we explain the fundamental characteristics using 
the Cramér-Rao bound in point estimation. 
Drawing an analogy from classical estimation theory, Helstrom formulated the quantum Cramér-Rao inequality using 
the Symmetric Logarithmic Derivative (SLD). 

The quantum estimator (corresponding to the estimation method) is determined by the SLD equation 
(Lyapunov equation) and is invariably a self-adjoint operator. This is equivalent to the fact that quantum measurement is described by $d\Pi(x)$.
Thus, its operational significance lies primarily in shaping a probability distribution rather than decision. 
 Therefore, it is appropriate to refer to it as a POVM from a measure-theoretic perspective.
 In this sense, quantum estimation operationally reduces to a Model B structure. Details will be discussed in the next section.

\begin{figure}
\centering{\includegraphics[width=7cm]{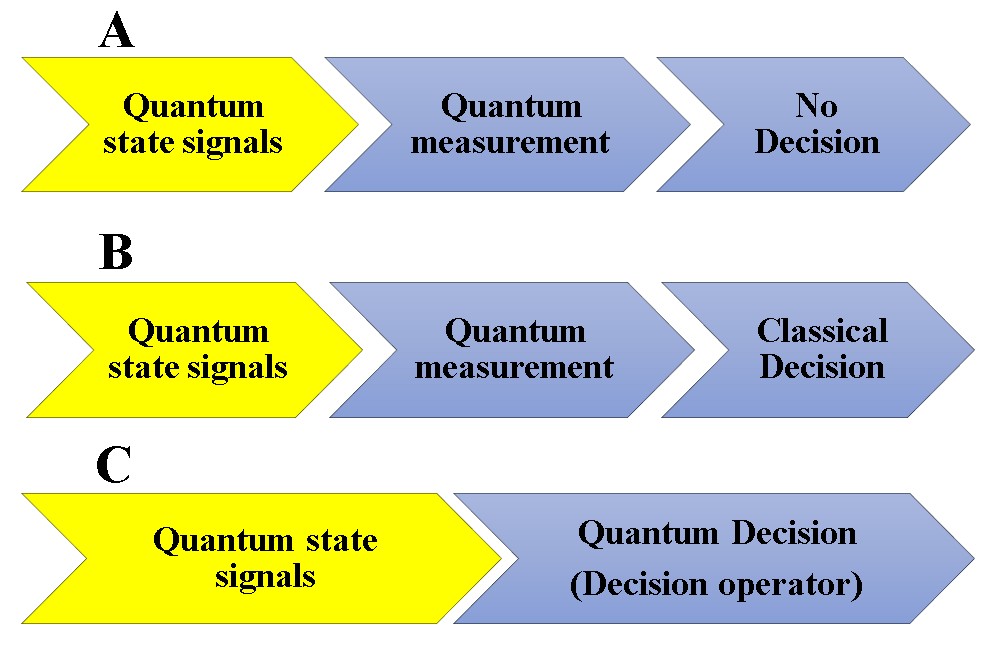}}
\caption{\textbf {Basic Structure of Quantum Measurement. Decision theory for discrete quantum signal ensembles falls under Type C, 
whereas estimation theory dealing with continuous quantities falls under Type B.}}
\end{figure}

\subsection{\textbf{Advanced Issues in Quantum Measurement}} \subsubsection{\textbf{Quantum Collective Measurement}} 
This section describes the theoretical structure of quantum measurement processes that perform collective measurements or collective decision operations on composite quantum state signals. \\
\textbf{(a) Discrete Systems}\\
 Let $\vert \Psi_{\mathbf{k}} \rangle$ be the $n$-fold tensor product quantum state representing a sequence of length $n$: 

\begin{equation} \vert \Psi_{\mathbf{k}} \rangle = \vert \psi_{k_1} \rangle \vert \psi_{k_2} \rangle \dots \vert \psi_{k_l} \rangle \dots \vert \psi_{k_n} \rangle,
 \end{equation} 

where $k_l \in \{1, 2, \dots, M\}$ for all $l \in \{1, 2, \dots, n\}$, and $\mathbf{k} = (k_1, k_2, \dots, k_n)$. Following Ban-Kurokawa-Hirota [46], 
the decision operator corresponding to the collective measurement can be defined as 
\begin{subequations} \begin{equation}
 \Pi_{\mathbf{j}} = \vert \Phi_{\mathbf{j}} \rangle \langle \Phi_{\mathbf{j}} \vert, \end{equation} \begin{equation} \vert
 \Phi_{\mathbf{j}} \rangle = \left( \sum_{\mathbf{k}} \vert \Psi_{\mathbf{k}} \rangle \langle \Psi_{\mathbf{k}} \vert \right)^{-1/2} \vert \Psi_{\mathbf{j}} \rangle,
  \end{equation} \begin{equation} P(\mathbf{j} \vert \mathbf{k}) = \mathrm{Tr} \rho_{\mathbf{k}} \Pi_{\mathbf{j}}, \end{equation}
   \end{subequations} 
where $\rho_{\mathbf{k}} = \vert \Psi_{\mathbf{k}} \rangle \langle \Psi_{\mathbf{k}} \vert$. 
This is referred to as a decision operator based on measurements utilizing quantum entanglement. 

In this framework, beyond the quantum effects arising from individual decision operations on each alphabet symbol,  
one can also harvest quantum advantages originating from multi-state entanglement or 
 non-orthogonality over the sequences.  To quantify such performance, it is essential to
  establish the nature of the quantum gain achieved through collective measurements under an appropriately specified evaluation function. 

Historically, the concept of quantum gain emerged in the context of the super-additivity of Shannon mutual information. Its ultimate characterization is 
established by the Holevo-Schumacher-Westmoreland (HSW) theorem [47, 48] concerning the classical capacity of quantum channels in discrete systems. 
Collective measurements play a crucial role in achieving this fundamental limit. Early concrete demonstrations of this phenomenon were presented in a series of
 works by Sasaki and Hirota [49] regarding super-additivity. However, the performance gains of collective measurements become particularly pronounced 
 when the alphabet is discrete and $M$-ary block-coded signals are employed [50]. 

Meanwhile, the general theory concerning the influence of collective quantum measurements on the quantum reliability function for finite-length codes was
 investigated by Burnashev and Holevo [51, 52]. Furthermore, the explicit quantitative advantage offered by collective measurements was rigorously 
 demonstrated in [46].\\
 \textbf{(b) Continuous Systems}\\
Describing collective measurements in systems with continuous alphabets is extremely challenging and requires further advanced mathematical tools. 
 However, it is possible to infer their effects from the perspective of channel capacity. The Holevo capacity for continuous systems with 
 coherent-state signals was established through a series of research papers by Holevo, Sohma, and Hirota [53, 54]. 
 While these results demonstrate a capacity exceeding that of classical systems, a detailed theoretical analysis regarding the specific quantitative effects 
 of collective measurements has not yet been fully conducted. However, it is shown by Sohma and Hirota that collective measurements are 
 actually not required to achieve the Holevo capacity in the regime of extremely weak received optical signals [55]. 
 
 Furthermore, regarding the theory of quantum reliability functions for continuous systems, Holevo et al. [56] proposed a conceptual definition; 
 however, their investigation only provides identifying the quantum cutoff rate for such continuous systems. \\
 \textbf{(c) Summary of section}\\ 
Collective measurements performed along the time axis inherently entail processing delays. Since modern information networks demand ultra-low latency, 
any quantum gain achieved through collective measurements must be substantial for the system to offer a practical advantage. In estimation theory as well, 
strict latency characteristics represent a critical operational requirement, necessitating the design of ultra-fast "one-shot" measurements. \\

\subsubsection{\textbf{Simultaneous Measurement of Non-Commuting Observables}} 
One type of quantum measurement closely tied to quantum estimation theory is the simultaneous measurement of non-commuting physical quantities. 
This section outlines the historical background of this topic. 
One of the most profound challenges in quantum measurement theory is formulating the simultaneous measurement of non-commuting observables. 
This issue has been scrutinized since the dawn of quantum mechanics and explored by numerous researchers. 

In the 1960s, approaches rooted in quantum optics garnered significant attention. Arthurs and Kelly [57], Gordon and Louisell [58], and She and Heffner [59]
 identified the presence of "excess quantum noise" associated with simultaneous measurements and developed a phenomenological description of this noise. 
 
 From a mathematical perspective, Holevo utilized Naimark's dilation theorem to represent the problem of measuring non-commuting observables 
 as a projection-valued measure (PVM) on an extended Hilbert space [31]. 
 This provided a unified framework for handling general quantum measurements—specifically, positive operator-valued measures (POVM) [60]. 
 Furthermore, pure mathematical research into more generalized quantum measurements was initiated by Davies [61], and ultimately, 
 the most important mathematical formulation was completed by Ozawa [62].
 
 Listed below are two theorems regarding the theoretical description of realization for quantum measurement in quantum optics.

\medskip \noindent \textbf{Theorem 1 (Naimark's  dilation Theorem):} \\ 
A positive operator-valued measure (POVM) on a Hilbert space $\mathcal{H}_S$ can be extended to an orthogonal resolution of the identity 
(projection-valued measure, PVM) on a larger Hilbert space. Conversely, projecting a PVM on an extended Hilbert space back onto 
$\mathcal{H}_S$ yields a POVM on $\mathcal{H}_S$. \\

In addition we may use the following theorem.

\medskip \noindent \textbf{Theorem 2 (Swap Operator Identity):} \\ 
Let $\rho_S(\alpha)=|\Psi(\alpha)\rangle\langle\Psi(\alpha)|$ be a signal Gaussian state on $\mathcal{H}_S$, and 
let $\rho_M(\beta)=|\Psi(\beta)\rangle\langle\Psi(\beta)|$ be a measurement-mode Gaussian state on $\mathcal{H}_M$. 
Here $\alpha, \beta$ are complex numbers.
Suppose these two independent modes form the tensor-product state $\rho_S(\alpha)\otimes \rho_M(\beta)$. 
By applying the SWAP operator $S_W$ to this two-mode system, the following identity holds:
\begin{equation}
\mathrm{Tr}[\rho_S \rho_M] = \mathrm{Tr}_{S,M} \left[ (\rho_S \otimes \rho_M) S_W \right].
\end{equation} 
Furthermore, the abstract probabilistic structure of complex amplitudes—originally defined via the inner product of two Gaussian states—
can be physically realized through projective measurements onto the symmetric and antisymmetric subspaces of the two-mode system. 

\medskip \noindent
To apply this fundamental framework to actual physical devices, a design theory for engineering the measurement state $\rho_M$ in Theorem~2 becomes
 indispensable. Reference [16] proposed incorporating a scheme-which controls the quantum fluctuations of 
 the complex amplitude of light—into the measurement system, directly embedding it into the optimal estimation framework. 
 The concrete formulation of this methodology is presented in Section~IV

\section{\textbf{Fundamentals of Quantum Estimation Theory}}
 In this section, we review the fundamental theory of the quantum Cram\'er-Rao inequality developed during the early period of quantum communication theory. 

\subsection{\textbf{Quantum Cram\'er-Rao Theory}} The quantum Cram\'er-Rao inequality encompasses four basic mathematical frameworks. 
Their essential concepts are outlined below. 

\textbf{(a) Helstrom's SLD Theory (1968)} [63]: \\ Consider a quantum state $\rho(\theta)$ that depends on an unknown single parameter $\theta$. 
The derivation of the quantum estimation process for estimating this parameter, along with the lower bound on the mean square error of the estimator, 
is formulated as follows (See appendix A): \\

$\textbf {Theorem 3}$ : 
The estimation bound for single parameter is given by following formula.
\begin{equation}
Var {\hat \theta} =Tr \rho ({\hat \theta} - {\theta})^2 \ge \frac{1}{Tr \rho(\theta) {L}_S^2} =\frac{1}{J_Q}
\end{equation}
where $L_S$ is called symmetric logarithm derivative (SLD), and is given by Lyapunov equation as follows (see apendix B):
\begin{equation}
\frac{\partial \rho(\theta)}{\partial \theta}=\frac{1}{2}[\rho(\theta){L}_S +{ L}_S \rho(\theta)]
\end{equation}
The equality in Eq (6) holds in the case as follows:
\begin{equation}
{L}_S=k(\theta)(\hat { \theta}-{ \theta})
\end{equation}
where $\hat {\theta}$ is estimator, $k(\theta)$ is a function of $\theta$, and the quantum estimator $\hat {\theta}$ is 
given by the self-adjoint operator corresponding to the parameter. Here, $J_Q$ denotes the Fisher information. \\

The derivation of inequality of Eq(6) employs the Cauchy-Schwarz inequality, reflecting the classical formulation; 
 thus, this method can be viewed as a direct quantum extension of classical systems. 

\textbf{(b) Yuen-Lax RLD Theory (1973)} [33]: \\In general, Helstrom's formulation of the SLD cannot 
handle the simultaneous estimation of non-commuting observables. To overcome this limitation, Yuen and Lax introduced the RLD
 (Right Logarithmic Derivative) method.
\begin{equation}
\frac{\partial \rho(\alpha)}{\partial \alpha}=L^{\dagger}  \rho(\alpha)
\end{equation}
where $\alpha$ is complex number. Yuen pointed out that the existence of the RLD relies on the Hilbert-Schmidt theory and the Riesz-Fréchet theorem. 
In a Hilbert space as Liouville space equipped with an inner product defined via the trace class, the differentiation operation is viewed as a linear map; 
the Riesz-Fréchet theorem then guarantees the existence of a non self-adjoint operator $L$ that represents this operation through the inner product.
In other words, one can say that $L^{\dagger}$ is algebraically determined using the Moore-Penrose pseudoinverse.

The lower bound in this Cramér-Rao inequality intrinsically incorporates the effects of the simultaneous measurement of non-commuting observables, 
making it suitable for explaining the excess quantum noise observed in standard optical heterodyne detection. However, a difficult problem arises 
regarding the physical realization of the operator $L$.
 
 \textbf{(c) Holevo Theory (1976--1978)} [31]: \\ This theory aims to provide a mathematically complete and general formulation of multi-parameter 
 estimation bounds that encompasses both Helstrom's and Yuen-Lax's theories. However, it assumes differentiability and linearity in the neighborhood 
 of the true parameter. While the quantum estimator is determined by the SLD, Holevo refined the quantum Cram\'er-Rao theory by incorporating 
 non-commuting effects and expressing the lower bound in terms of commutation relations. 
 
 Consequently, the resulting lower bound consists of  the SLD lower bound plus an additional term depending on the commutation relations. 
 Such a generalized theory gives rise to a new challenge:  finding physical systems that achieve this bound. 
 This is known as the realization problem. Two primary methodologies have been pursued to address this challenge: the communication-theoretic approach [16] 
 and the statistical approach based on quantum local asymptotic normality(Quantum LAN) [64] (See Fig.3). 
 
 \textbf{(d) Measurement Process Separation Theory (1977)} [16]: \\In practical applications, the problem of realization becomes important. 
 In oreder to proceed the problem, a theory that avoids the kind of mathematical complexity mentioned above is required. 
 A methodology was proposed to decouple the quantum estimation problem into  a signal description space and a measurement space. 
 The signal description space specifies the estimator, whereas the measurement space describes  the quantum measurement system including non-commuting parameter measurements (Fig.4).
  
 In this approach, the estimator is first determined using  the SLD in the signal description space, and the Cram\'er-Rao bound is subsequently reduced 
 to a problem of ideal simultaneous measurement of  that estimator.  A distinct feature of this approach is the introduction of degrees of freedom to 
 control quantum noise in the quantum measurement process. Since the present work adopts this theory, the details will be elaborated 
 in the next section and the appendix A and B.

\begin{figure}
\centering{\includegraphics[width=7cm]{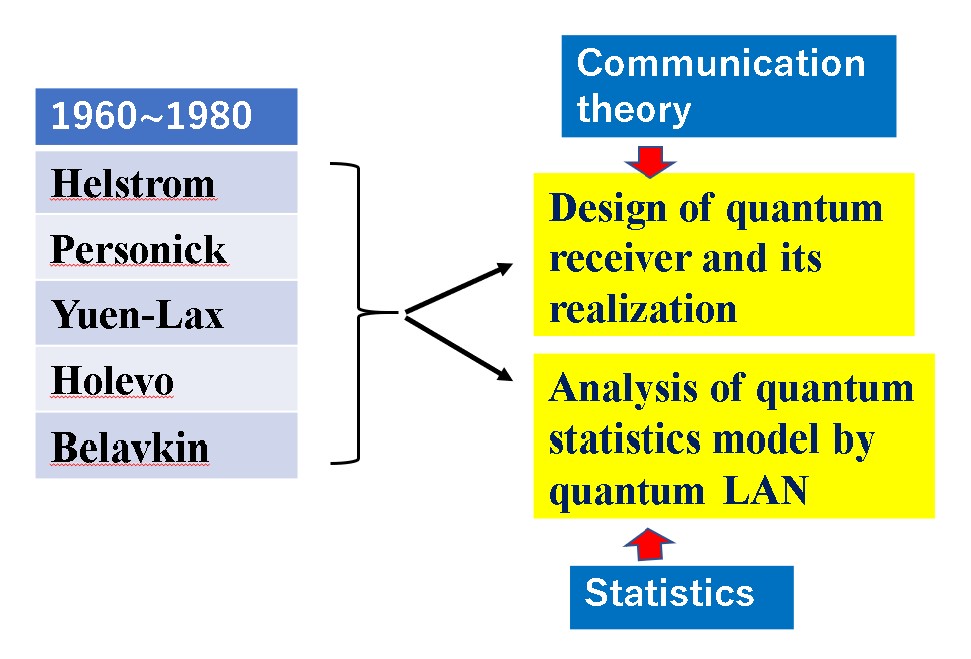}}
\caption{\textbf {Current concepts of quantum estimation theory based on communication theory and statistics for exploring quantum advantage.}}
\end{figure}

\begin{figure}
\centering{\includegraphics[width=8cm]{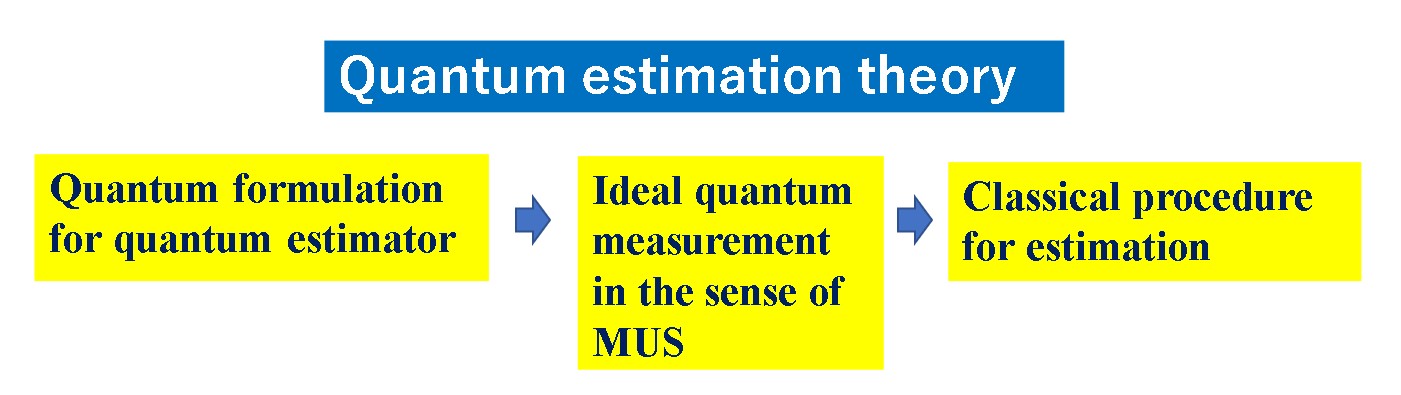}}
\caption{\textbf {Measurement process separation theory in quantum estimation therory for estimation  including 
non-commuting physical quantities as parameters.The degrees of freedom for quantum noise control in quantum measurement are encompassed 
by MUS (Minimum Unceratainty State). Classical estimation methods are applied after the measurement.}}
\end{figure}

\subsection{\textbf {Quantum MMSE theory}}
 In standard estimation problems within communication theory, The signal is a random process. In other words, the parameter is a modulated signal, 
 not merely an unknown parameter. In that case, the source's prior probability distribution is assumed to be known, and 
 the Minimum Mean Square Error (MMSE) criterion is typically used to evaluate estimation performance in such models. 

 A quantum version of such an estimation method was formulated by  Personick in 1971 [32]. In this case as well, it is difficult to specify 
 the correct estimation bound if the quantum estimators are non-commuting observables. 
 Similar to the Cramer-Rao inequality, methodologies to resolve this issue are required.

\subsection{\textbf {Summary of section}}
From the explanations so far, these mark the conclusion of the "First Generation" of research.
The second generation can be characterized as the pursuit of real-world examples that achieve quantum advantage, based on such theories. 
Numerous researchers have already yielded valuable results in this endeavor. To further advance these achievements, 
we discuss the methodology proposed by the authors below.\\

\section{\textbf {Theoretical Framework for   Quantum Noise Control in Estimation Theory}}

This section provides a detailed explanation of the theory of quantum estimation based on the separate theory (Fig.4) [16] 
for a general quantum Gaussian state [65].

\subsection{\textbf{Formulation of separation theory}}
The concept is very simple. 
When considering quantum estimation problems, the signal representation space ${\cal H}_s$ 
and the measurement space ${\cal H}_{qm}$ are separated.
Quantum estimation formulas are formulated in signal space. Once quantum estimators are determined, the corresponding ideal measurements are 
considered in the measurement space. Describing this as a protocol makes implementation possible.

\textbf {Protocol:}\\
Step 1: Derive the quantum estimators  for simultaneous estimation of two parameters $ x_c, x_s$ based on the SLD equations.
\begin{subequations}
\begin{equation}
\frac{\partial \rho_S(x_c)}{\partial x_c}=\frac{1}{2}[\rho_S(x_c){ L}_S +{L}_S\rho_S(x_c)] 
\end{equation}
\begin{equation}
\frac{\partial \rho_S(x_s)}{\partial x_s}=\frac{1}{2}[\rho_S(x_s){L}_S +{L}_S \rho_S(x_s)]
\end{equation}
\end{subequations}
Step 2: Construct a measurement projector that expresses simultaneous measurement of noncommuting quantities 
based on the minimum uncertain state for the two operators obtained.\\

In the first step, if the quantum estimators are not non-commuting, the first step provides the correct bound.
When these are non commuting,  the following shows how to derive a generalized quantum measurement 
that represents their simultaneous measurement. including the quantum noise control.

\subsection{\textbf{Mathematical formulation for generalized measurment process with quantum noise control }}

We have been formulating the theory of quantum state control and simultaneous measurement of non-commuting quantities based on Lie algebra [6].
Here we introduce it and its application to the formulation of estimation theory. \\

\textbf {Definition 3} : 
An algebra is called a Lie algebra when its multiplication satisfies the following two conditions:
\begin{subequations}
\begin{equation}
[{\bf A}, {\bf A}]=0, 
\end{equation}
\begin{equation}
[[{\bf A}, {\bf B}], {\bf C}] +[[{\bf B}, {\bf C}], {\bf A}] + [[{\bf C}, {\bf A}], {\bf B}]=0
\end{equation}
\end{subequations}
where $[\quad ]$ is the commutation relation.\\

An example of its application in quantum mechanics is the Heizenberg-Weyl group: $W_1$. 
The unitary representation of the sub group of $W_1$ is 
\begin{equation}
{\bf T}({\bf g})=\exp\{is\}\exp\{\alpha{\bf a}^{\dagger} -\alpha^*{\bf a}\}
\end{equation}
where ${\bf g}$ is the elements of Lie algebra.\\

\textbf {Definition 4} : 
For any element $|\psi >$ of Hilbert space, the quantum state constructed by the following formula 
\begin{equation}
{\bf T}({\bf g})|\psi >=|\alpha_{\psi} >_G
\end{equation}
is called generalized coherent state.\\

Here let us consider non-compact groups in Lie algebras.
With respect to Hermite forms, a set that constitute linear isometric groups is non-compact group.
SU(1,1) is a typical example.

The Lie algebra of SU(1,1) has the following elements :${\bf K}_1, {\bf K}_2, {\bf K}_0$. Their commutation relations are
\begin{subequations}
\begin{equation}
[{\bf K}_1, {\bf K}_2]=-i{\bf K}_0, [{\bf K}_2, {\bf K}_0]=i{\bf K}_1
\end{equation}
\begin{equation}
[{\bf K}_0, {\bf K}_1]=i{\bf K}_2
\end{equation}
\end{subequations}
Then we define ${\bf K}_{\pm}={\bf K}_1 \pm i{\bf K}_2$.
Then we have 
\begin{equation}
[K_0, K_{\pm}]=\pm K_{\pm}, \quad [K_2, K_{\pm}]=2K_0
\end{equation}
Let us define the Casimir operator of $\{K_{j}\}$ as follows:
\begin{eqnarray}
C_2 &=&K_0^2 -K_1^2 -K_2^2 =K_0^2 -\frac{1}{2}(K_+K_- + K_-K_+)\nonumber \\
&=&\tau(\tau - 1)I
\end{eqnarray}
The unitary representation of SU(1,1) is determined using the value of $\tau$ as an indicator.
The unitary representaion of SU(1,1) is given by 
\begin{equation}
{\bf  T}^{\tau}({\bf g})=\exp\{\zeta {\bf K}_+ - \zeta^* {\bf K}_-\}
\end{equation}
 The following state is also the generalized coherent state.
\begin{equation}
{\bf T}^{\tau}({\bf g})|\psi > =|\alpha_{\tau} >_G
\end{equation}
where $|\alpha_{\tau} >$ is called a base state.
When the above state satisfies the minimum uncertainty relation for two observales, 
it is called  the ideal generalized coherent state.\\

Here for the quadrature amplitude of light wave, the elements of SU(1,1) are given as follows:.
\begin{subequations}
\begin{equation}
{\bf K}_+=\frac{1}{2}{{\bf a}^{\dagger}}^2,  {\bf K}_-=\frac{1}{2} {\bf a}^2, 
\end{equation}
\begin{equation}
 {\bf K}_0=\frac{1}{4}({\bf a}{\bf a}^{\dagger} + {\bf a}^{\dagger}{\bf a})
 \end{equation}
\end{subequations}
Then we have 
\begin{eqnarray}
&& {\bf T}^{\tau}({\bf g})=\exp\{\frac{z}{2}({\bf a}^2 - {{\bf a}^{\dagger}}^2)\} \\
&&{\bf T}^{\tau}({\bf g}){\bf a} {{\bf T}^{\tau}}^{\dagger}({\bf g})=\mu {\bf a} + \nu {\bf a}^{\dagger} \equiv {\bf b}\\
&& |\alpha_{\tau} > = | 0 >, \quad |\mu|^2 - |\nu|^2=1  \nonumber
\end{eqnarray}
Then the following state is the ideal generalized coherent state.
\begin{eqnarray}
&&{\bf b} |\beta, \mu, \nu > = \beta |\beta, \mu, \nu > \\
&&{\bf T}^{\tau}({\bf g}) |0 > = |0, \mu, \nu >
\end{eqnarray}
where $\beta = \mu \alpha + \nu \alpha^*=x_{c(M)} + i x_{s(M)}$. 
The above state is the minimum uncertainty state for $x_{c(M)}$ and $x_{s(M)}$ when $\mu$ and $\nu$ are the real as 
\begin{equation}
\mu=\frac{\lambda +1}{\sqrt {2\lambda +1}}, \quad \nu= \frac{\lambda}{\sqrt {2\lambda +1}}
\end{equation}
where $ 0 \le \lambda \le \infty $. Thus, we have
\begin{subequations}
\begin{equation}
\Delta x_{c(M)}^2  =\frac{1}{4} |\mu - \nu|^2 =\frac{1}{4}( \frac{1}{2\lambda +1})=\frac{1}{4}\frac{1}{T_{qm}} 
\end{equation}
\begin{equation}
\Delta x_{s(M)}^2 =\frac{1}{4} |\mu + \nu|^2= \frac{1}{4}(2\lambda +1)=\frac{1}{4} T_{qm}    
\end{equation}
\begin{equation}
\Delta x_{c(M)}^2 \times \Delta x_{s(M)}^2 = \frac{1}{16}
\end{equation}
\end{subequations}
where $\lambda$ is a control parameter, and $T_{qm}=2\lambda +1$. Thus it is a minimum uncertainty state.
Then the measurement projector for the step 2 is defined  as follows:
\begin{equation}
d\Pi(x_{c(M)}, x_{s(M)}) = |\beta, \mu, \nu ><\beta, \mu, \nu |d^2\beta/\pi
\end{equation}
This is called the generalized heterodyne POVM.
The final measurement probablity density is given by 
\begin{equation}
P(x_c, x_s)=Tr \rho_S d\Pi (x_{c(M)}, x_{s(M)})
\end{equation}
The above non orthogonal projector provides a control of the balance of excessive noise due to simultaneous measurement by 
the parameter $\lambda$. The specific form of Eq (27) was presented in Reference [16] and Appendix.
This scheme is called ``{\it a generalized heterodyne receiver}".
Let us show how to use the above formula in the following. \\

\subsection{\textbf{Basis for estimation bound of complex amplitude by generalized heterodyne receiver}}
\subsubsection{\textbf{Coherent state}}
The signal state is given as follows:
\begin{equation}
|\alpha >, \quad \alpha={\bar x}_c + i {\bar x}_s=Ae^{i\theta}
\end{equation}
First, the solutions of quantum estimater are given by SLD equation as follows:
\begin{subequations}
\begin{equation}
 L_{S1}=4(\frac{{\bf a} + {\bf a}^\dagger}{2} - {\bar x_c}), 
 \end{equation}
 \begin{equation}
 L_{S2}= 4(\frac{{\bf a} - {\bf a}^\dagger}{2i} - {\bar x_s})
 \end{equation}
\end{subequations}
Thus, quantum estimators correspond to  $X_C$ and $X_S$ of the optical mode.
The estimation bounds are given as the variances for the following output distribution.
\begin{equation}
P(x_c, x_s)=Tr |\alpha ><\alpha |  d\Pi (x_{c(M)}, x_{s(M)})
\end{equation}
That is, we have the following bounds.
\begin{subequations}
\begin{equation}
Var {\hat x}_c =Var(SLD) + Var(M)= \frac{1}{4} +\frac{1}{4}\frac{1}{T_{qm}}
\end{equation}
\begin{equation}
Var {\hat x}_c = Var(SLD) + Var(M)=\frac{1}{4} +\frac{1}{4}T_{qm}
\end{equation}
\end{subequations}
where $Var(SLD)$  and $Var(M)$ are variances of SLD on the description space and the quantum measurement for 
simultaneous measurement of non communtative signals, respectively.

Then the total minimum bound is 
\begin{equation}
\sigma^2_T=\min_{\lambda} (Var {\hat x}_c + Var {\hat x}_s )= 1, \quad \lambda=0
\end{equation}
This is equivarent to Yuen-Lax bound.\\

\subsubsection{\textbf{Squeezed state}}
A squeezed state in which the coherent components and variance are independent can be described as follows:
\begin{equation}
|\alpha; \xi>= D(\alpha) S(\xi) |0 >
\end{equation}
where $S(\xi)$ , $\xi =ze^{-2i\phi}$ is a squeezing operator with squeezing parameter $z_s$, and $D(\alpha)$ is a shift operator. 
When the signal state is the above with $\phi =0$ or $\pi/2$, the bounds are given by 
\begin{subequations}
\begin{equation}
Var {\hat x}_c =\frac{1}{4}e^{\mp 2z_s} +  \frac{1}{4} \frac{1}{T_{qm}}
\end{equation}
\begin{equation}
Var {\hat x}_s =\frac{1}{4}e^{\pm 2z_s} + \frac{1}{4}T_{qm}  
\end{equation}
\end{subequations}
These provide the basis for the following analysis.\\

Finally, we will summarize the basis of  generalized heterodyne POVM. 
SU(1,1) was introduced to define equivalence classes of minimum uncertainty states. This theoretical framework is best suited for defining equivalence classes of minimum uncertainty states based on complex amplitudes, rather than on momentum and coordinate systems. The generalized heterodyne POVM is defined within this theory. Furthermore, to effectively apply separation theory to automatic control systems, it is necessary to utilize Theorem 2 or related theorems. In particular, while the generalized heterodyne POVM yields measurement probabilities in the form of the squared inner product, the actual implementation of this in a physical system requires invoking Theorem 2.

\section{\textbf{Basis for estimation bound of optical phase}}
In recent years, the application of quantum estimation theory to optical phase estimation has attracted significant attention. 
In this section, we will introduce the problem posed by Monras et al which is arguably the most important among them. \\

\subsection{\textbf {Formulae of phase estimation for general Gaussian noise}}
\subsubsection{\textbf{When the parameters depend on the mean vector }}
Let $x_c + ix_s=Ae^{i\theta}$ be the complex amplitudes of the received signals. Here, the amplitude may be either known or unknown.
Let us assume that the signals are disturbed by Gaussian noise with unequal variances $\sigma_c, \sigma_s$.
\begin {equation}
\vec{ Y}=\begin{bmatrix}x_c\\ x_s \end{bmatrix}+n=A \begin{bmatrix}\cos \theta \\ \sin \theta \end{bmatrix} +n
\end{equation}
where $n \sim N(0, \Sigma_V)$:
\begin{equation}
\Sigma_V=\begin{bmatrix} \sigma^2_c & r\sigma_c \sigma_s \\ r\sigma_c \sigma_s & \sigma^2_s \end{bmatrix}
\end{equation}
where $r$ is a correlation coeffcient.
Let us define the Fisher information matrix.
\begin{equation}
J= \begin{bmatrix} J_{AA} & J_{A\theta} \\ J_{\theta A} & J_{\theta \theta} \end{bmatrix} 
\end{equation}
Based on the measurement of the complex amplitude, the final estimation bound of the phase is given by 
\begin{equation}
Var {\hat \theta} \ge \frac{J_{AA}}{J_{\theta \theta} J_{A A} - J^2_{A \theta}}
\end{equation}

Let us denote the inverse of the matrix $\Sigma_V$ as follows:
\begin{equation}
\Sigma^{-1}_V = \begin{bmatrix} a & c \\ c & b \end{bmatrix}
\end{equation}
Then conponents of the Fisher information matrix are given as follows:
\begin{subequations}
\begin{equation}
J_{AA}= a \cos^2 \theta + 2cs  \sin \theta  \cos \theta + b \sin^2 \theta 
\end{equation}
\begin{equation}
J_{\theta \theta}=A^2 (a  \sin^2 \theta - 2cs \sin \theta \cos \theta + b \cos^2 \theta 
\end{equation}
\begin{equation}
J_{A  \theta}= A[(b-a)  \sin \theta  \cos \theta + c(\cos^2 \theta - \sin^2 \theta)]
\end{equation}
\end{subequations}
where 
\begin{subequations}
\begin{equation}
a =\frac{1}{\sigma^2_c (1-r^2)}
\end{equation}
\begin{equation}
b= \frac{1}{\sigma^2_s (1-r^2)} 
\end{equation}
\begin{equation}
c= - \frac{r}{\sigma_c \sigma_s (1- r^2)}
\end{equation}
\end{subequations}

\subsubsection{\textbf{When the parameter depends only on the covariance}}
If the parameter depends only on the covariance, the score function becomes a quadratic form.
In this case, the phase change is represented as a rotation of the covariance matrix. Consequently, the parameter can be treated as a single quantity.

The Fisher information for estimating the phase corresponding to that rotation, based on the simultaneous measurement of complex amplitudes,
 is given by the following.
\begin{equation}
J(\theta) = \frac{1}{2} Tr[(\Sigma^{-1}_V \frac{\partial \Sigma_V}{\partial \theta})^2]
\end{equation}

\subsection{\textbf {Basis for quantum Gaussian noise}}
In the theory of quantum measurement separation, the quantum estimator is determined as the solution to the SLD equation, and the quantum measurement corresponds to the associated ideal measurement problem. Since the quantum estimators are $X_c$ and $X_s$ based on the preceding discussion, the ultimate estimation bound is analyzed in terms of the properties of the measurement probability distribution obtained through the simultaneous measurement of these two quantities.

\subsubsection{ \textbf{Coherent state signal}}
In the initial stage, if there is no reference phase, it is necessary to simultaneously measure the complex amplitude of the optical mode 
 in order to estimate the phase of the optical signal. The complex amplitude coresponds to non commuting observable.

Here, we assume that the quantum state is a coherent state.
Let us  set the receiver with $\lambda = 0$  or $T_{qm}=1$ which corresponds to the conventional heterodyne receiver. 
Then we can use the post-measurement processing based on the probability density with the variance of Eq(32). The log-likelihood function is given by 
\begin{equation}
\ln (P(y_c, y_s|\theta))= \frac{A}{\sigma^2_T/2}(y_c \cos(\theta) +y_s \sin(\theta))
\end{equation}
where $y_c=x_c + n$, and $y_s=x_s + n$.
Then the classical ML estimator is  ${\hat \theta} = tan^{-1}(y_c/y_s)$, and the Fisher information is
\begin{equation}
 J(\theta) =\frac{A^2}{\sigma^2_T/2}
 \end{equation}
Here if polar coordinate representation is adopted such as $y_c=R\cos \Phi, \quad  y_s = R\sin \Phi$, 
we have the conventional joint probability density for $R$ and $\Phi$.
When we use an aproximation such as $R\cong A$, $\Phi \cong \theta$, the ML estimate is efficient. That is
\begin{equation}
Var {\hat \theta} =E[(\Phi - \theta)^2] \cong \frac{1}{ J^(\theta) }= \frac{\sigma^2_T/2}{A^2}=\frac{1}{SNR}
\end{equation}
While such standard systems offer no quantum advantage, the characteristic quantum benefits associated with squeezed states
—as described in the following sections—have been demonstrated. 
This paper introduces these concepts and discusses methodologies for realizing their potential benefits.

\subsubsection{\textbf{Bound of Monras-Olivares-Paris}}

In the previous subsection, we showed the phase estimation formula when the variances of both components of the complex amplitude are equal.
On the other hand, when the light source is in a squeezed vacuum state with $\phi=\pi/2$, the coherent component is zero and the variances of 
each component of the complex amplitude are different. The quantum state with phase shift as a signal for the squeezed vacuum state is 
described by $U(\theta)$ as follows:
\begin{equation}
U(\theta) S(\xi)|0><0| S(\xi)^{\dagger}U^{\dagger}(\theta)
\end{equation}

Therefore, the coherent component is not necessary.to formulate the estimation problem of optical phase.
In this case, it is possible to estimate the rotation of the ellipse axes by exploiting the anisotropy of the quantum noise variance. 

Monras and  Olivares-Paris have discussed the important properties of the estimation bound of optical phase by using the squeezed vacuum state.
They concluded that the ideal value for the phase estimation bound is as follows.

\begin{subequations}
\begin{equation}
\Delta \theta^2  \times \Delta G^2 \ge \frac{1}{F}
\end{equation}
\begin{equation}
\Delta G^2=\frac{1}{2} \sinh^2 (2z)
\end{equation}
\end{subequations}
where $F$ is a scaling constant which depends on the definition. Thus, the phase estimation bound is 
\begin{equation}
Var {\hat \theta}= \frac{1}{F/2\sinh^2 2z}\cong  \frac{F}{2} e^{-4z}, \quad z \gg 1, \quad \forall \theta
\end{equation}
This is called Monras- Olivares-Paris bound.

In general, it is important to discuss the methodology for realizing the aforementioned bound.
This becomes possible only through the measurement of the complex amplitude of light. They investigated various optical reception schemes. 
The most fundamental results are as follows.

Here one can consider the homodyne receiver to obtain the practical scheme of the estimation of $\theta$.
Olivares-Paris  investigated the use of a homodyne receiver as a specific method to achieve the Monras bound.
\begin{subequations}
\begin{equation}
Var {\hat \theta} = \frac{8({\sigma^2_{\theta}})^2}{\sinh^2 2z \sin^2  2\theta} 
\end{equation}
\begin{equation}
\sigma^2_{\theta}=\frac{1}{4}[ e^{-2z} \cos^2 \theta + e^{2z} \sin^2 \theta]
 \end{equation}
 \end{subequations}

When $z=-\frac{1}{2}\log(tan(\theta))$, the above is 
\begin{equation}
Var {\hat \theta} = \frac{1}{2\sinh^2 2z} = \frac{1}{2} \tan^2 2\theta,
\end{equation}
Thus, to bring Eq(49a) closer to the form of Eq(48), strict conditions on $z$ or the signal phase are necessary.

On the other hand,  if the conventional heterodyne is adopted, they showed that  the estimation bound is given as follows:
\begin{equation}
Var({\hat \theta})  =\frac{1}{4\sinh^2 z} \cong  e^{-2z}, \quad  : z \gg 1, \quad  0 \le  \theta <\pi
\end{equation}
Also they discussed the detailed comparison between homodyne and heterodyne measurements [36].\\

\section{\textbf{Performance based on generalized heterodyne POVM}}
In this section, we apply the generalized heterodyne POVM to the issue of optical phase estimation.

\subsection{\textbf{Signal independent noise control receiver}}
When a generalized heterodyne receiver is adopted, the conditional probability density is given by two dimensional Gauss distribution based on 
\begin{equation}
p(x_c, x_s|\theta) = |<0| S(\xi)^{\dagger}U^{\dagger}(\theta)|\beta: \mu,\nu >|^2/\pi
\end{equation}
The total covariance matrix at the initial state  is given by 

 \begin{eqnarray}
\Sigma_T(\theta=0)&=& \Sigma_T(0)=\Sigma_S(0) + \Sigma_M \nonumber \\
&=&\begin{bmatrix} \sigma^2_{sq(c )}&0 \\ 0 & \sigma^2_{sq(s )}\end{bmatrix}+\frac{1}{4}\begin{bmatrix} T_{qm} 
&0 \\ 0& \frac{1}{T_{qm}}\end{bmatrix}
\end{eqnarray}

 Then the effect of phase rotation is given by
 \begin{equation}
\Sigma_T(\theta)=U(\phi (\theta)) \Lambda  U(\phi (\theta))^T
\end{equation}
 where $\Lambda$ is the diagonal matrix based on the eigenvalues $\tau_{\pm}(\theta)$ of the total covariance matrix, and where
 \begin{subequations}
 \begin{equation}
\sigma_{sq(c)}^2 = \frac{1}{4}e^{\pm 2z_s} 
\end{equation}
\begin{equation}
\sigma^2_{sq(s)}=\frac{1}{4}e^{\mp 2z_s}  
\end{equation}
\begin{equation}
T_{qm}=2\lambda +1
\end{equation}
\end{subequations}

The formula of Fisher information becomes as follows.
\begin{eqnarray}
J(\theta)&=&E[\{\frac{\partial}{\partial \theta} \log p(x_c, x_s|\theta)\}^2] \nonumber \\
&=&\frac{1}{2} Tr \{\Sigma^{-1}_T(\theta) \frac{\partial \Sigma_T(\theta)}{\partial \theta}\}^2 \nonumber \\
&=& J_{eigen} + J_{rot}\nonumber \\
&=& \frac{1}{2}[(\frac{\tau_{+}'}{\tau_{+}})^2 +(\frac{\tau_{-}'}{\tau_{-}})^2]+\frac{(\tau_{+} - \tau_{-})^2}{\tau_{+} \tau_{-}} (\theta ')^2
\end{eqnarray}
Then, let us describe the total covariance matrix as follows:
\begin{equation}
\Sigma_T(\theta)= \begin{bmatrix} a(\theta) &  b(\theta) \\ b(\theta) & c(\theta) \end{bmatrix}
\end{equation}
where
\begin{subequations}
\begin{equation}
a(\theta)=\sigma^2_{sq(c)} \cos^2  \theta + \sigma^2_{sq(s)}\sin \theta+ \frac{1}{4}T_{qm} 
\end{equation}
\begin{equation}
c(\theta)=\sigma^2_{sq(c)}\sin^2  \theta + \sigma^2_{sq(s)} \cos^2  \theta  +\frac{1}{4T_{qm}}
\end{equation}
\begin{equation}
b(\theta )= (\sigma^2_{sq(c )}- \sigma^2_{sq(s)} )\sin \theta \cos \theta
\end{equation}
\end{subequations}
The eigenvalue of the above matrix is 
\begin{equation}
\tau_{\pm}(\theta)=\frac{a(\theta)+c(\theta)}{2} \pm \frac{1}{2} \sqrt{(a(\theta)-c(\theta))^2 + 4 b(\theta)^2}
\end{equation}
Then, let us show the final form.
\begin{equation}
J(\theta)= \frac{4\delta_s^2 \delta_n^2  (q^2+r^2) \sin^2(2\theta) }{r^2(q^2 - r^2) }+
\frac{4\delta_s^2 ( \delta_s + \delta_n \cos(2\theta))^2}{r^2(q^2 - r^2) }
\end{equation}
where 
\begin{subequations}
\begin{equation}
\delta_s=(\sigma_{sq(c)}^2 - \sigma_{sq(s)}^2)/2, \quad \delta_n=(T_{qm} -1/T_{qm})/8
\end{equation}
\begin{equation}
q=(\sigma_{sq(c)}^2 + \sigma_{sq(s)}^2)/2 + (T_{qm} +1/T_{qm})/8
\end{equation}
\begin{equation}
r^2=(a(\theta)-c(\theta))^2 + 4b(\theta)^2
\end{equation}
\end{subequations}
Thus, the performance strongly depends on the signal phase. 
When  $\cos 2\theta \sim 0$, we have 
\begin{equation}
J(\theta) \sim  \frac{(\sigma^2_{sq(c )}- \sigma^2_{sq(s)})^2}{(\sigma_{sq(c)}^2 + 1/4T_{qm})(\sigma_{sq(s)}^2 + 1/(4T_{qm}))}
\end{equation}
Here $r^2$ with $\theta \sim 0$ in the denominator and $(\delta_s + \delta_n)^2$ with $\theta \sim 0$ in the numerator become equal, casing the measurement noise in the numerator to cancel out.

When the control parameter satisfies the following relations, 
\begin{equation}
\frac{1}{4} T_{qm} \sigma^2_{sq(s)} =1/16, \quad \frac{1}{4T_{qm}} \sigma^2_{sq(c)}=1/16
\end{equation}
we have 
\begin{equation}
Var ({\hat \theta}) = \frac{1}{J(\theta, T_{qm})} \cong 4e^{-4z}, \quad \theta \sim 0
\end{equation}
When the conventional heterodyne $T_{qm}=1$ gives for  phase $ 0 \le  \theta < \pi$  as follows:
\begin{equation}
\tau_+(\theta) \tau_-(\theta) =(\sigma^2_{sq(c )}+\frac{1}{4})(\sigma^2_{sq(s)} + \frac{1}{4})
\end{equation}
Then 
\begin{equation}
Var ({\hat \theta}) =\frac{1}{J(\theta, T_{qm}=1) } \cong  e^{-2z}
\end{equation}
This is equivalent to Eq(51).

Thus, these results represent the performance achieved when the quantum state control in the generalized heterodyne is independent of the signals.

\begin{figure}
\centering{\includegraphics[width=7cm]{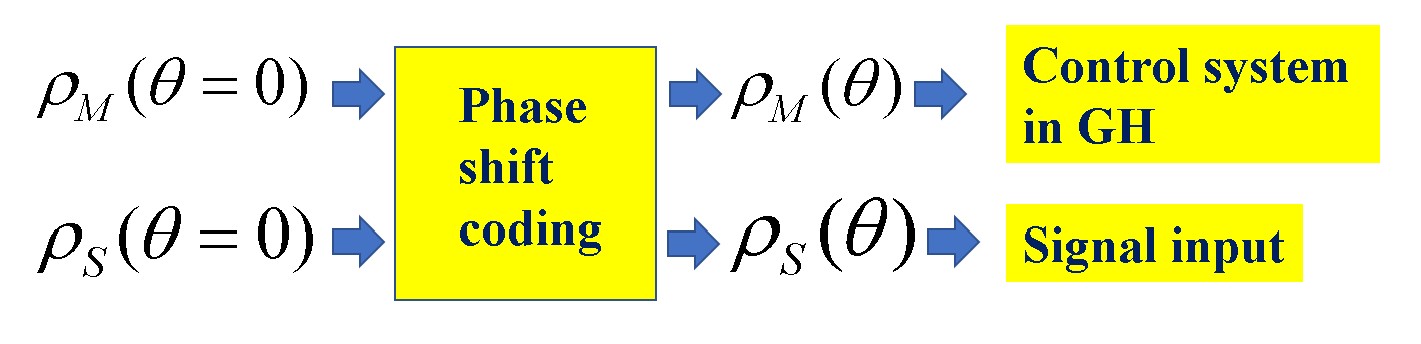}}
\caption{\textbf {Framework for auto quantum noise control. GH is the generalized heterodyne. This is an application of a technique
 that realizes the projection of signal states via generalized POVMs in a two-mode format. However, the quantum states of the two modes employed therein 
 are encoded simultaneously.}}
\end{figure}

\subsection{\textbf{Auto-signal dependent control system}}
In the previous section, we analyzed the case where quantum noise control is independent of the signal. The results showed 
that the Monras-Olivares-Paris bound can be achieved only when the signal phase takes a specific value. Here, we present a method to achieve this 
without dependence on the signal phase, similar to conventional heterodyne detection.

As shown in the section IV, the generalized heterodyne POVM allows for the control of the balance of quantum noise characteristics and 
the orientation of the ellipse axes through the manipulation of the quantum state of the ancillary system from the theorem 2.
Here we consider the quantum version of auto receiver parameter control in the conventional communication technology by using 
the generalized heterodyne POVM (see Appendix C)

Let us set the system such that the auxiliary quantum state for realizing generalized heterodyne quantum noise control is supplied through 
the same channel as the signal (Fig.5 and appendix C). This enables automatic and adaptive control of the noise characteristics of the measurement process, 
even without signal measurement. In this case, we have
\begin{subequations}
\begin{equation}
U(\theta) \rho_S U(\theta)^T \otimes U(\theta) \rho_M U(\theta)^T
\end{equation}
 \begin{equation}
\Sigma_T(\theta)= U(\theta) (\Sigma_S(0) +\Sigma_M)U(\theta)^T 
\end{equation}
\end{subequations}
The eigenvalues of the total covariance becomes independent of $\theta$.
Then the product of eigenvalues  is 
\begin{equation}
\tau_+(\theta) \tau_-(\theta)=(\sigma^2_{sq(c)}+ T_{qm}/4 )(\sigma^2_{sq(s)} + 1/4T_{qm})
\end{equation}
Since the eigenvalues are independent of the phase, the Fisher information is given  by only rotation of  axes as follows:
\begin{equation}
J(\theta) =  \frac{((\sigma^2_{sq(c)} +1/4T_{qm})- (\sigma^2_{sq(s)} + 1/(4T_{qm}))^2}
{(\sigma^2_{sq(c)}+ T_{qm} /4)(\sigma^2_{sq(s)} + 1/(4T_{qm}))}
\end{equation}

Let us denote the specifications for the required auxiliary quantum state. The energy of the squeezed state of the signal is determined by 
the squeezing parameter. That is $<n>=|\sinh(2z)|^2$. 
In this case, one should prepare an auxiliary quantum state that is also a squeezed state characterized by the same squeezing parameter. 
Hence, we can assume that the control parameters are $T_{qm}/4=\sigma_{sq(c)}^2$, and $1/(4T_{qm})=\sigma_{sq(s)}^2$. 

Specifically, this is structured as follows: 
Both the signal and the auxiliary quantum states serve as inputs to the same modulation system.
It is treated as a single shot measurement setup, and they achieve the following bound.
\begin{equation}
Var ({\hat \theta})   \cong e^{-4z}, \quad \forall \theta \in [0, \pi)
\end{equation}

Thus, we obtain a very simple formula for  Monras-Olivares-Paris bound of optical phase estimation.
This result implies that the performance of the bound  is achieved due to  the covariances of the anisotropic noise in the signal and 
in the measurement system work together to enhance phase sensitivity. 
This can be attributed to the synergistic effect of dual anisotropic noise in a system configured using generalized heterodyning that involves 
synchronized rotation of two quantum states. The advantages of the system adopted here are discussed in Appendix D.

\section{ \textbf{Quantum MMSE estimation for analog modulation tracking}}

This section aims to present examples of applying squeezed states and generalized heterodyne measurements to these problems.
In systems employing squeezed states, the relative phase between the signal and the measurement system is generally fixed. 
However, the system addressed in this paper operates on the premise that the amplitude and phase of the probe quantum state are random process. 
Such a scenario necessitates a theoretical framework based on the simultaneous measurement of the complex amplitude of the probe light; in other words,
 heterodyne theory is fundamental. Below, we examine a system designed to estimate changes in the coherent component (utilizing a squeezed coherent state) 
 based on the aforementioned model. Furthermore, to achieve higher speeds, the estimation device relies on a "one-shot" decision process.

\subsection{ \textbf{Formulation of MMSE}}
The mean square error (MSE) is one of the principal measures of the quality of an estimator. 
The quantum Bayes MSE is defined as follows:
\begin{equation}
C({\hat \Theta})=\int p(\theta)\{Tr[\rho_{\theta}({\hat \Theta} - \theta I)^2]\}d\theta
\end{equation}
where $\Theta$ is a self adjont operator and the measurement of $\Theta$ gives the outcome of which is an estimate of the value $\theta$. 
The formulation for this kind of estimation was given by Personick as follows [32]. \\

\textbf{ Theorem-4}\\
If we make the definitions,
\begin{subequations}
\begin{equation}
\Gamma = \int  p(\theta)  \rho_{\theta} d\theta, 
\end{equation}
\begin{equation}
\gamma = \int \theta p(\theta) \rho_{\theta} d\theta
\end{equation}
\end{subequations}
the MMSE estimator must satisfy
\begin{equation}
\Gamma {\hat \Theta}_{opt} + {\hat \Theta}_{opt} \Gamma = 2 \gamma
\end{equation}

This formalization provides the estimeter in the description space; that is, it is the same as the aforementioned SLD formalization.
If the estimation operators are commuting, then it directly provides the properties of the final estimation limit.

\subsection{ \textbf{Estimation bound for complex amplitude}}

In communication systems, complex amplitude—or phase and amplitude—serves as the information carrier.
In this section, the limitations of simultaneous estimation of complex amplitudes in optical tracking are shown.
For simplicity, we will only consider quantum noise. A priori density function of the parameter is given by 
\begin{equation}
\rho(\alpha) = \int Gauss(\alpha, {\bar \alpha}, \Sigma_{\alpha})|\psi (\alpha)><\psi (\alpha)| d^2 \alpha
\end{equation}
The elements of covariance matrix $\Sigma_{\alpha}$ are $\sigma_{\alpha(c)}=\sigma_{\alpha(s)}$.
From Theorem 4, the two quantum estimators can be determined separately.
\begin{subequations}
\begin{equation}
{\hat X_c}=k({\bar \alpha}, \sigma_{\alpha})[(a +a^{\dagger})/2] 
\end{equation}
\begin{equation}
{\hat X_s} =k({\bar \alpha}, \sigma_{\alpha})[(a -a^{\dagger})/2i]
\end{equation}
\end{subequations}
These are the non commuting. Here, we adopts the separate theory described in the section IV, and introduce the simultaneous measurement of 
two observables determined by Personick formula and apply classical estimation methods to the variables in the post-measurement.\\

Since the analysis of systems employing generalized heterodyne POVMs is complex, the analysis here is conducted using the following limited model.

Here, the signal vector of $\alpha=x_c +i x_s=A e^{i\theta}$ is
 \begin{equation}
\vec{X}=  \begin{bmatrix} x_c \\ x_s \end{bmatrix}
\end{equation}
The channel mode  of the signal is a linear estimation as follows:
 \begin{equation}
\vec{Y}=  \begin{bmatrix} y_c \\ y_s \end{bmatrix} = w\begin{bmatrix} x_c \\ x_s \end{bmatrix}+ \begin{bmatrix} n_c \\ n_s \end{bmatrix}
\end{equation}
The  a priori distribution is assumed as Gaussian distribution with zero mean and variance $\sigma^2_{\alpha}$.
\begin{equation}
p(x_c, x_s): {\cal N}(0, \sigma^2_{\alpha})
\end{equation}
The quantum state is assumed as squeezed state.\\

\subsubsection{ \textbf{System of  generalized heterodyne without auto control}}
First, we consider a system that does not have an automatic synchronization mechanism with the signal.

The covariance of the total noise by squeezed state and quantum measurement depends  only on the phase as follows:
\begin{equation}
\Sigma_{total}(\theta)= U(\theta) \Sigma_{sq}(0) U(\theta)^T + \Sigma_{qm}(0)
\end{equation}
where $\theta$ is the angle of the complex amplitude and squeezed axis, and 
\begin{subequations}
 \begin{equation}
\Sigma_{sq(c)}=  \begin{bmatrix} \sigma^2_{sq(c)} & 0 \\ 0 & \sigma^2_{sq(s)} \end{bmatrix}, 
\end{equation}
\begin{equation}
 \Sigma_{qm}(0)=  \begin{bmatrix} \sigma^2_{qm(c )}& 0 \\ 0 & \sigma^2_{qm(s )}\end{bmatrix}
\end{equation}
\end{subequations}

Here we adopt ``Partial or local  linearization" method. First, the phase is fixed. Then, the problem goes to a linear Gauss estimation.
The posterior probability distribution is indeed a Gaussian distribution and has the covariance $\Sigma_{post}$, and its inverse matrix is 
\begin{equation}
\Sigma_{post}^{-1}(\theta)= \Sigma_{\alpha}^{-1} + \Sigma_{total}^{-1}(\theta)
\end{equation}
The estimater is 
\begin{equation}
{\hat \alpha}=[\sigma^{-2}_{\alpha}I + \Sigma^{-1}_{total}(\theta)]^{-1}\Sigma^{-1}_{total} \vec {Y}
\end{equation}
where $\theta$ is fixed.

Let us assume that the signal to noise ratis is high. 
\begin{subequations}
\begin{equation}
SNR_{x_c} = \frac{A^2}{\sigma^2_{sq(c)}+\sigma^2_{qm(c)}} \gg 1
\end{equation}
\begin{equation}
SNR_{x_s} = \frac{A^2}{\sigma^2_{sq(s)}+\sigma^2_{qm(s)}} \gg 1
\end{equation}
\end{subequations}
Then we can use partial linearization. 
That is, 
\begin{equation}
\theta \sim {\hat \theta}= arg(y_c + iy_s)
\end{equation}
Then, the estimator can be expressed as follows:
\begin{equation}
{\hat \alpha}=\sigma^2_{\alpha}I[\sigma^{2}_{\alpha}I + \Sigma_{total}({\hat \theta})]^{-1} \vec {Y}
\end{equation}

Under a high signal-to-noise ratio, it becomes as follows.
\begin{subequations}
\begin{equation}
C({\hat x_c})\cong \frac{\sigma^2_{\alpha}[ \sigma^2_{\alpha} ( \sigma^2_{sq(c)}+ \sigma^2_{sq(s)})/2 +\sigma^2_{qm(c)}) + \Delta}
{\sigma^4_{\alpha} + \sigma^2_{\alpha} Z + \Delta}
\end{equation}

\begin{equation}
C({\hat x_c})\cong \frac{\sigma^2_{\alpha}[ \sigma^2_{\alpha} ( \sigma^2_{sq(c)}+ \sigma^2_{sq(s)})/2 +\sigma^2_{qm(s)}) + \Delta}
{\sigma^4_{\alpha} + \sigma^2_{\alpha} Z + \Delta}
\end{equation}
\end{subequations}
where 
$Z=\sigma^2_{sq(c)} + \sigma^2_{sq(s)} + \sigma^2_{qm(c)} + \sigma^2_{qm(s)}$, 
$\Delta=1/8 + \sigma^2_{sq(c)}\sigma^2_{qm(s)} + \sigma^2_{sq(s)}\sigma^2_{qm(c)}$. 
Then, we have 
\begin{equation}
C({\hat x_c}) -C({\hat x_s})= \frac{\sigma^4_{\alpha} (\sigma^2_{qm(c)} -\sigma^2_{qm(s)})}
{\sigma^4_{\alpha} + \sigma^2_{\alpha} Z + \Delta}
\end{equation}

An intuitive approximation is as follows.
\begin{equation}
C({\hat x_c}) \sim \frac{1}{4} e^{2z},  \quad  C({\hat x_s})\sim \frac{1}{8} e^{2z}
\end{equation}
where $\sigma^2_{sq(c)}=1/4e^{2z}$, $\sigma^2_{sq(s)}=1/4 e^{-2z}$, $\sigma^2_{qm(c)}=1/4 T_{qm}$,  $\sigma^2_{qm(s)}=1/(4T_{qm})$,
$\sigma^2_{sq(c)}=\sigma^2_{qm(c)}, \sigma^2_{sq(s)}=\sigma^2_{qm(s)}$,  
$\sigma^2_{sq(c)}\gg \sigma^2_{sq(s)}, \sigma^2_{qm(c)}\gg \sigma^2_{qm(s)}$.  \\

These are attributable to the rotation of anisotropic noise. As a result, coherent states and standard heterodyne detection are optimal 
when the measurement axis of the receiver is fixed. It means that there is no quantum advantage due to squeezed state.\\

\subsubsection{ \textbf{Single system with auto-signal dependent control system}}

Next, a system that automatically controls measurement quantum noise is adopted (see Appendix C).  
This is the same system described in Section VI-B.
According to the auto control of the $\rho_M$, 
the covariance of the total noise by squeezed state and quantum measurement is 
\begin{equation}
\Sigma_{total}(\theta)= U(\theta)\{ \Sigma_{sq}(0)  + \Sigma_{qm}(0)\}U(\theta)^T
\end{equation}

 In this setup, the observed value is a linear function of the signal, and its mean corresponds to the signal value. 
 Meanwhile, the variance is determined by the sum of the anisotropic variances of the signal system and the measurement system; 
 although the covariance matrix rotates depending on the signal, the magnitude of this variance remains independent of the signal. 
 Consequently, the problem can be reduced to a linear Gaussian model.

First, the posterior covariance for the random signal obtained through measurement is expressed as follows:
\begin{equation}
\Sigma_{X|Y}{\theta}=(\sigma^{-2}_{\alpha} I + \Sigma^{-1}_{total}(\theta))^{-1}
\end{equation}

Given that the prior probability is isotropic and the estimator possesses a rotationally covariant structure
—meaning the statistics remain invariant when the input and output are rotated by the same amount—averaging yields:
\begin{equation}
\Sigma_{X|Y} = c I
\end{equation}
In other words, the MMSE becomes isotropic.

However, we can exploit the property that the eivenvalues are fixed and that the variances of two components of noise rotate in tandem.
To create anisotropy, we switch to a system that utilizes only one of the two components, as shown below:
\begin{equation}
y_c=x_c + N
\end{equation}
The other component is discarded. Since $x_c$ is observed via a scalar linear Gaussian channel,
 the estimated evaluation function becomes independent of phase rotation, thereby matching the standard formula without approximation.

 In this case, the system  exhibits the following characteristics under the  $\sigma^2_{sq(c)} =\sigma^2_{qm(c)} = 1/4e^{-2z} \ll 1$ and 
 $\sigma^2_{\alpha} \gg \sigma^2_{sq(c)}$:
\begin{subequations}
\begin{eqnarray}
C({\hat x_c}) &=& \frac{\sigma^2_{\alpha}(\sigma^2_{sq(c)} + \sigma^2_{qm(c)})}{\sigma^2_{\alpha} +\sigma^2_{qm(c)}} \nonumber \\
&\cong&  \frac{1}{2} e^{-2z} \\
C({\hat x_s})&=& \sigma^2_{\alpha}
\end{eqnarray}
\end{subequations}

For comparison, let us assume a noise-free homodyne setup in which the receiver detects one component of the complex amplitude. 
In this homodyne mechanism, 
since the phase cannot be set in advance, so phase synchronization is required. Consequently, various approaches must be considered.
Thus, due to the rotation of anisotropic noise, the evaluation function becomes dependent on the noise component with the larger variance.
\begin{subequations}
\begin{equation}
C({\hat x_c}) \sim \frac{1}{4} e^{2z}, \quad C({\hat x_s})\sim \sigma^2_{\alpha}
\end{equation}
or
\begin{equation}
C({\hat x_s}) \sim \frac{1}{4} e^{2z}, \quad C({\hat x_c}) \sim \sigma^2_{\alpha}
\end{equation}
\end{subequations}
The latter terms of the above equations  correspond to the component that is not measured. 
Thus, the quantum advantage due to squeezed state disappears.
\\

\subsubsection{ \textbf{Twin systems with auto-signal dependent control system}}
Twin-modes redundancy is one of the environmentally responsive technologies in communication theory. 
This section will examine its quantum version (Fig.6).
Here, we prepare two instances of the system structure described in the previous section (System A and System B) and 
operate them within a single modulation system.

A twin squeezed state system is defined as follows:a two  squeezed coherent state $\rho_A \otimes \rho_B$  is prepared at the prove source such that 
the squeezing directions of their variances are opposite to each other. 
The receiver employs a generalized heterodyne scheme to measure the twin signals separately, with the balance of their quantum state control parameters 
set to have reciprocal characteristics.

The coherent components (complex amplitude) of this pair are simultaneously encoded by a single modulator and transmitted into 
the communication channel without loss; in other words, the coherent components of the twin squeezed state carry identical information. 

\begin{subequations}
\begin{equation}
C({\hat x_c})_A=\sigma^2_{\alpha}, \quad  C({\hat x_s})_A = \frac{1}{4} e^{-2z}
\end{equation}
and
\begin{equation}
C({\hat x_c})_B = \frac{1}{4} e^{-2z}, \quad C({\hat x_s}) _B= \sigma^2_{\alpha}
\end{equation}
\end{subequations}
This performace cannot be attained by any conventional two homodyne schemes or heterodyne measurement.
This gain results from the combined effect of introducing redundancy and controlling quantum noise; it does not transcend fundamental limits.
That is, it corresponds to the quantum version of a communication technology that utilizes redundancy through the simultaneous modulation of two modes.

\begin{figure}
\centering{\includegraphics[width=8cm]{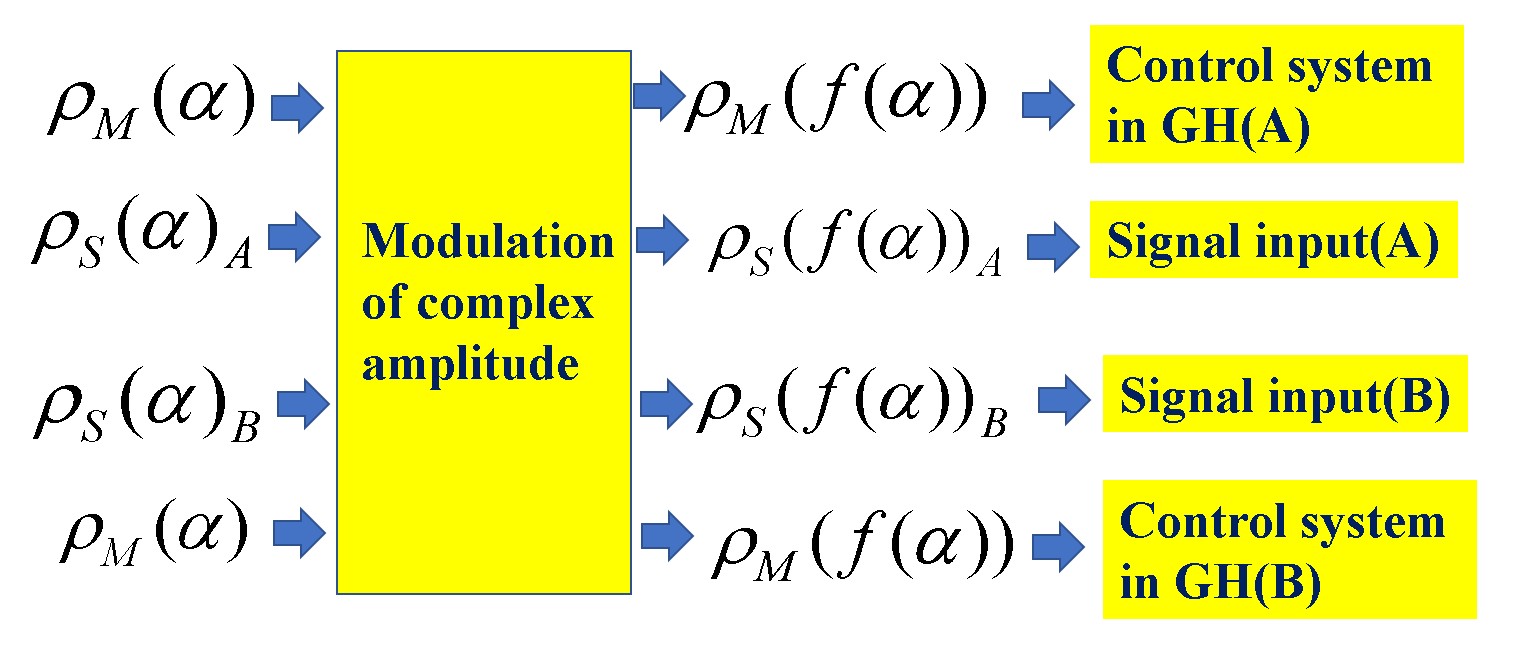}}
\caption{\textbf {Framework for twin systems with auto quantum noise control. GH is the generalized heterodyne.
The scheme shown in Figure 5 is configured as a twin system, with the four modes sharing the same modulation information. 
This is analogous to a classical technique that enhances reception performance through redundancy.}}
\end{figure}

\subsection{ \textbf{Transformation to amplitude and phase estimation from complex amplitude of twin system}}
In this section, we consider the mechanism for converting  the complex amplitude estimation system 
of the twin system discussed in the previous section into estimates of the amplitude and phase.
In the twin system described in the previous section, the modulation signals applied to both systems are identical,
 and outputs $x_c$ and $x_s$ are obtained from System A and System B, respectively. Thus, although the setup involves the combination of two channels, 
 it is mathematically equivalent to performing simultaneous complex amplitude estimation within a single system. 
In this twin system, the modes of the two channels are independent, resulting in uncorrelated noise components. 
The conversion process is carried out below under these conditions.

This constitutes a nonlinear coordinate transformation for a bivariate Gaussian distribution of complex amplitudes $\alpha= A e^{i\theta}$.
\begin{equation}
A={\sqrt {x_{c}^2 + x_{s}^2}}, \quad \theta = \tan^{-1}(\frac{x_{s}}{x_{c}})
\end{equation}
When the first-stage complex amplitude estimation circuit of the twin system involves independent Gaussian errors with variance 
for the complex amplitude as follows:
\begin{subequations}
\begin{equation}
 C({\hat x_s})_A = \frac{1}{4} e^{-2z} \ll 1
\end{equation}
and
\begin{equation}
C({\hat x_c})_B = \frac{1}{4} e^{-2z} \ll 1
\end{equation}
\end{subequations}
Here, the off-diagonal elements of the covariance matrix for the two systems are zero, because the noise in the two systems is independent.
In this case, the conversion to amplitude and phase follows a relatively simple formula. Thta is, it means SNR $\gg 1$. 
Thua we can adopt that 
the probability distribution of the amplitude is a Rice distribution with the variance $\sigma_{out}$ and 
the phase distribution is a wrapped distribution, as shown below.
\begin{equation}
p(\theta)=\frac{1}{\sqrt {2\pi} \sigma_{out}}\Sigma_{k=-\infty}^{\infty}\exp[-\frac{(\theta - {\bar \alpha} + 2k\pi)^2}{2\sigma_{out}}]
\end{equation}
As a result, the estimation bounds for amplitude and phase are 
\begin{subequations}
\begin{equation}
Var (A) \cong  \frac{1}{4} e^{-2z}
\end{equation}
\begin{equation}
Var (\theta) \cong  \frac{1}{4A^2} e^{-2z}
\end{equation}
\end{subequations}
The twin system provides high SNR for the complex amptude estimation. 
So when one wishes to understand the correspondence regarding amplitude and phase estimation, a very simple relationship
—as described above—is obtained..\\

\section{ \textbf{Conclusion}}
In this paper, we presented a method for addressing the problem of estimating non-commutative parameters by applying a theory that treats quantum estimators 
and the measurement problem separately. This framework integrates SLD (Symmetric Logarithmic Derivative) theory—formulated for quantum estimators within 
the quantum signal space—with the ideal simultaneous measurement of the corresponding estimators. 
We demonstrated the application of this framework to estimation problems in systems utilizing Gaussian states, such as squeezed states, as probe sources.
 Specifically, by applying generalized heterodyne theory and the swap operator identity theorem, we showed that the Monras-Paris bound can be achieved 
through one-shot measurement. This approach employed an analogy to the concept of differntial phase encoding (or Manchester codeing) used 
in classical systems. 

Furthermore, we adopted these ideas as a methodology to extend Personick's analog modulation system to a two-variable case. 
The results revealed that combining twin parallel redundancy of quantum states with quantum noise control leads to the existence of useful estimation bounds
 in delay-free one-shot estimation.

 However, the specific construction method for the generalized heterodyne measurement—
which plays an essential role in these results—remains undetermined.
We have identified two tasks for future work. The first is the construction of a unified theory for realizing generalized POVMs, 
and the second is an investigation into whether the proposed method can be extended to sequential adaptive estimation [67, 68, 69]. 
We plan to address these issues in a subsequent paper, drawing upon Personick's theory [32, 66].

\section*{\textbf{Acknowledgement}}
The author would like to thank F.Futami, K. Kato, K.Tanizawa, G. Masada, K.Nakahira and T.Usuda for their enthusiastic participation in discussions. 
Additionally, AI (Gemini, Copilot, Chat GPT of free version) was used to improve the English phrasing.

\section*{\textbf{Appendix}}
\renewcommand{\theequation}
{A-\arabic{equation}}
\setcounter{equation}{0}

\section*{\textbf{A: Mathematical Meaning of the Lyapunov Equation in Quantum Estimation Theory}}
\label{app:lyapunov-sld}

In this appendix, following Holevo[31], we clarify the role of the Lyapunov-type equation that defines the symmetric logarithmic derivative (SLD). 
For simplicity, we restrict attention to a smooth one-parameter family $\rho(\theta)$ of positive-definite density operators on 
a finite-dimensional Hilbert space $\mathcal{H}$ of dimension $d$. This restriction avoids both the support-related nonuniqueness of the SLD for 
rank-deficient states and the domain issues associated with unbounded operators in infinite-dimensional Hilbert spaces. 

Although the state model is restricted to the finite-dimensional, full-rank case, the outcome space of the POVM is not assumed to be finite;
continuous outcome spaces are allowed. For a more general treatment beyond these assumptions, we refer the reader to Holevo [31].

\textbf{1. Lyapunov Equation Defining the SLD}
Let $\rho(\theta)>0$ and ${\rm Tr}\rho(\theta)=1$. The derivative of the state,
\begin{equation}
\dot{\rho}_{\theta}:=\frac{d\rho(\theta)}{d\theta},
\end{equation}
satisfies
\begin{equation}
\dot{\rho}_{\theta}^{\dagger}=\dot{\rho}_{\theta},\qquad{\rm Tr}\dot{\rho}_{\theta}=0.
\end{equation}
The SLD $L_{\theta}^{S}$ is defined by the Lyapunov equation
\begin{equation}
\dot{\rho}_{\theta}=\frac12\left(\rho(\theta)L_{\theta}^{S}+L_{\theta}^{S}\rho(\theta)\right).\label{eq:appendix_sld}
\end{equation}
For a classical probability model $p_{\theta}(x)$, the score function
$\ell_{\theta}(x)=d\log p_{\theta}(x)/d\theta$ satisfies
\begin{equation}
\frac{dp_{\theta}(x)}{d\theta}=p_{\theta}(x)\ell_{\theta}(x).
\end{equation}
Thus, Eq.~\eqref{eq:appendix_sld} may be regarded as a quantum counterpart of this relation, in which the product is symmetrized 
because the density operator and the SLD need not commute.
Taking the trace of both sides of Eq.~\eqref{eq:appendix_sld} gives
\begin{equation}
{\rm Tr}\rho(\theta)L_{\theta}^{S}=0.\label{eq:appendix_sld_zero}
\end{equation}
Furthermore, expanding $\rho(\theta)$ in an eigenbasis as
\begin{equation}
\rho(\theta)=\sum_ip_i(\theta)|i\rangle\langle i|,\qquad p_i(\theta)>0,
\end{equation}
we obtain
\begin{equation}
\left\{L_{\theta}^{S}\right\}_{ij}=\frac{2\left\{\dot{\rho}_{\theta}\right\}_{ij}}{p_i(\theta)+p_j(\theta)}.\label{eq:appendix_sld_element}
\end{equation}
Since $p_i(\theta)+p_j(\theta)>0$, the SLD is uniquely determined. Moreover, since $\dot{\rho}_{\theta}$ is self-adjoint, 
$L_{\theta}^{S}$ is also self-adjoint.
For self-adjoint operators $A$ and $B$, define
\begin{equation}
\left\langle A,B\right\rangle_{\rho}^{S}=\frac12{\rm Tr}\rho(AB+BA).\label{eq:appendix_sld_inner}
\end{equation}
For $\rho>0$, this defines an inner product on the real vector space of self-adjoint operators. 
The SLD quantum Fisher information is then defined by
\begin{equation}
J_Q(\theta)=
\left\langle
L_{\theta}^{S},
L_{\theta}^{S}
\right\rangle_{\rho(\theta)}^{S}
=
{\rm Tr}\rho(\theta)(L_{\theta}^{S})^2
=
{\rm Tr}\dot{\rho}_{\theta}L_{\theta}^{S}.
\label{eq:appendix_qfi}
\end{equation}

\textbf{2. SLD Quantum Cram\'er--Rao Inequality and Estimator Operator}
Let $M(d\hat{\theta})$ be a real-valued POVM on
$\mathbb{R}$ with finite second moment.
The probability distribution of the outcome is
\begin{equation}
\mu_{\theta}(d\hat{\theta})
=
{\rm Tr}
\rho(\theta)M(d\hat{\theta}).
\end{equation}
Its mean and variance are defined by
\begin{equation}
E_{\theta}\{M\}
=
\int
\hat{\theta}\,
\mu_{\theta}(d\hat{\theta}),
\end{equation}
and
\begin{equation}
D_{\theta}\{M\}
=
\int
\left(
\hat{\theta}-E_{\theta}\{M\}
\right)^2
\mu_{\theta}(d\hat{\theta}).
\end{equation}
The measurement is said to be locally unbiased at a reference point $\theta_0$ if
\begin{equation}
E_{\theta_0}\{M\}
=
\theta_0,
\qquad
\left.
\frac{d}{d\theta}
E_{\theta}\{M\}
\right|_{\theta=\theta_0}
=
1.
\label{eq:appendix_local_unbiased}
\end{equation}

Define the first- and second-moment operators by
\begin{equation}
X_M
:=
\int
\hat{\theta}\,
M(d\hat{\theta}),
\qquad
X_M^{(2)}
:=
\int
\hat{\theta}^{2}\,
M(d\hat{\theta}),
\label{eq:moment-operators-short}
\end{equation}
and let
\begin{equation}
Y_{\theta}
=
X_M-E_{\theta}\{M\}I.
\end{equation}

Since $X_{M}^{(2)}-X_{M}^{2}\geq0$,
\begin{eqnarray}
D_{\theta}\{M\}&=&\left\langle Y_{\theta},Y_{\theta}\right\rangle_{\rho(\theta)}^{S}+
\operatorname{Tr}\left\{\rho(\theta)\left(X_{M}^{(2)}-X_{M}^{2}\right)\right\}\nonumber \\
&\geq& \left\langle Y_{\theta},Y_{\theta}\right\rangle_{\rho(\theta)}^{S}.
\label{eq:variance-bound-short}
\end{eqnarray}
Moreover, Eqs.~\eqref{eq:appendix_sld} and \eqref{eq:appendix_sld_zero} yield
\begin{equation}
\frac{\mathrm{d}}{\mathrm{d}\theta}E_{\theta}\{M\}
=
\left\langle L_{\theta}^{S},Y_{\theta}\right\rangle_{\rho(\theta)}^{S}.
\label{eq:mean-derivative-short}
\end{equation}
Therefore, the Cauchy--Schwarz inequality gives
\begin{equation}
D_{\theta}\{M\}J_Q(\theta)
\geq
\left\{\frac{\mathrm{d}}{\mathrm{d}\theta}E_{\theta}\{M\}\right\}^{2}.
\label{eq:holevo-bound-short}
\end{equation}
Following Holevo [31], imposing the local unbiasedness
condition \eqref{eq:appendix_local_unbiased} yields the SLD quantum
Cram\'er--Rao bound
\begin{equation}
D_{\theta_0}\{M\}
\geq
\frac{1}{J_Q(\theta_0)}.
\label{eq:qcr-bound-short}
\end{equation}


The lower bound in Eq.~\eqref{eq:qcr-bound-short} can be attained at a fixed reference point $\theta_0$. Assuming $J_Q(\theta_0)>0$, define
\begin{equation}
A_{\theta_0}^{\ast}
:=
\theta_0I
+
\frac{1}{J_Q(\theta_0)}L_{\theta_0}^{S}.
\label{eq:optimal-estimator-short}
\end{equation}
Let $M^{\ast}$ be the spectral measure of $A_{\theta_0}^{\ast}$, and use its measurement outcomes as estimates of $\theta$. Since
$X_{M^{\ast}}=A_{\theta_0}^{\ast}$,
Eqs.~\eqref{eq:appendix_sld_zero},
\eqref{eq:appendix_qfi}, and \eqref{eq:mean-derivative-short} give
\begin{equation}
E_{\theta_0}\{M^{\ast}\}
=
\theta_0,
\qquad
\left.
\frac{\mathrm{d}}{\mathrm{d}\theta}
E_{\theta}\{M^{\ast}\}
\right|_{\theta=\theta_0}
=
1.
\label{eq:local-unbiased-optimal-short}
\end{equation}
Thus, $M^{\ast}$ is locally unbiased at $\theta_0$.
Furthermore, because $M^{\ast}$ is the spectral measure of $A_{\theta_0}^{\ast}$,
\begin{equation}
\begin{aligned}
D_{\theta_0}\{M^{\ast}\}
&=
\operatorname{Tr}
\rho(\theta_0)
\left(
A_{\theta_0}^{\ast}-\theta_0I
\right)^2\\
&=
\frac{1}{J_Q(\theta_0)^2}
\operatorname{Tr}
\rho(\theta_0)
\left(
L_{\theta_0}^{S}
\right)^2
=
\frac{1}{J_Q(\theta_0)}.
\end{aligned}
\label{eq:optimal-variance-short}
\end{equation}
Hence, the lower bound in Eq.~\eqref{eq:qcr-bound-short} is attained.

In summary, for a finite-dimensional family of positive-definite quantum states, the Lyapunov equation provides a unique correspondence 
between the derivative of the state and the SLD, and the SLD quantum Fisher information determines a lower bound on the variance of 
locally unbiased estimators. Moreover, at a fixed reference point $\theta_0$, a measurement attaining this bound can be constructed from the SLD. 
This construction is the quantum analogue of the equality
case of the classical Cram\'er--Rao inequality.


In the complex-amplitude estimation considered in Section~IV, this one-parameter construction is applied separately to the two parameter directions $x_c$ 
and $x_s$. It determines the corresponding estimator operators for the two quadrature components.

 Although each estimator operator can individually be associated with an optimal spectral measurement, the two estimator operators do not commute.


Consequently, their spectral measurements do not admit an exact joint measurement. Their joint measurement is therefore treated as a generalized measurement in the measurement space in Section~IV, where a generalized heterodyne POVM based on a minimum-uncertainty state is introduced.

\begin{figure}[htbp] \centering \includegraphics[width=8cm]{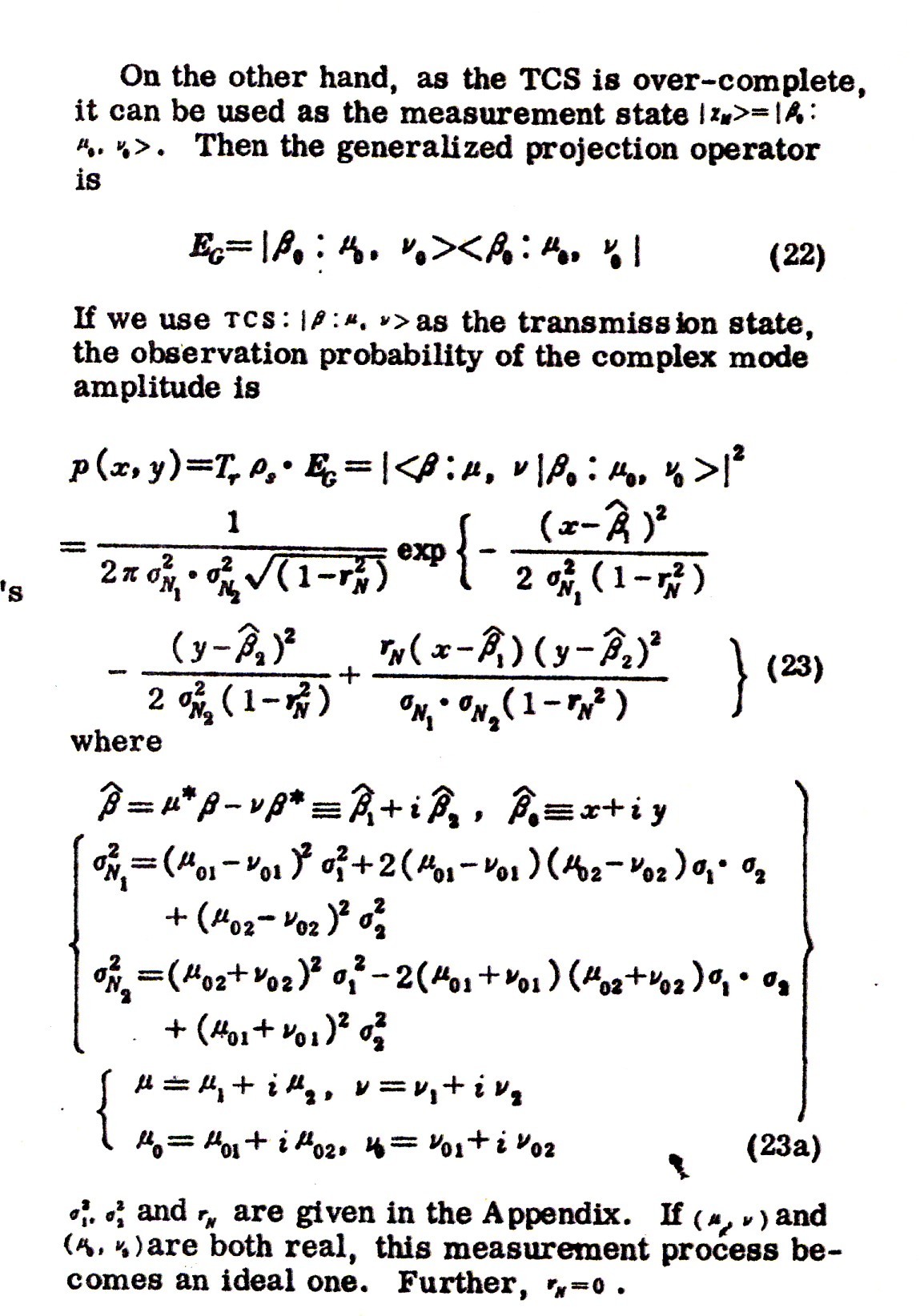}
 \caption{\textbf{The original idea was first published in Japanese in 1977 [16] 
and was also made available internationally through the English translation published in \textit{Electronics and Communications in Japan}. 
This figure is a reproduction from the English translation published in 1977 by Scripta Publishing Co.}} 
\label{fig:fig5} \end{figure}

\section*{\textbf{B: Background of the Measurement Process Separation Theory}}

This section explains the historical background of the theoretical framework adopted in this study. 
The mathematical foundation of SLD was described in Appendix A; there, it was shown that SLD is defined by a Lyapunov equation for the quantum state
 (or density operator) corresponding to the signal, and that its solution automatically determines both the quantum estimator and the quantum measurement 
 (as an orthogonal projection). As noted there, since this solution directly prescribes the quantum measurement, it is impossible to derive exact limits 
 when the quantum estimators are non-commutative. To circumvent this issue, theories such as Right Logarithmic Derivative (RLD) and Holevo theory 
 were proposed.

While Holevo theory provides the most general formulation, it inevitably gives rise to new challenges during the implementation phase. 
Various approaches have been proposed to address these implementation issues. One such approach utilizes methods from statistical asymptotic theory; 
however, since quantum systems are inherently physical systems, the application of asymptotic theory implicitly presupposes infinite energy or time delays. 
In optical communication systems and signal processing, delays associated with signal processing cannot be tolerated. 
Therefore, to ensure efficient estimation that realistically approaches estimation bounds, the development of non-asymptotic estimation methods is 
desirable, and numerous proposals have already been made.

Prior to those studies, Hirota proposed the "quantum measurement separation method"—which applies SLD theory to the problem of estimating 
non-commuting observables—in a series of papers on noise control in quantum optics published between 1975 and 1977. 
In optical signals, the fundamental physical quantity carrying information is the complex amplitude, which is a non-commuting observable; consequently, 
the estimation of these quantities became a subject of research.

 Optimizing the measurement process in the "quantum measurement separation method" 
involves the theory of "Minimum Uncertainty States (MUS)" associated with the physical quantity in question.
By introducing equivalence classes of these MUSs into the measurement system, it becomes possible to control quantum noise through a paradigm 
of quantum state control that differs from conventional approaches.

The standard Gaussian quantum state is the coherent state, which exhibits equal uncertainty in both components. Equivalence classes of coherent states 
are represented by two-photon coherent states (TCS) [13, 70]. This concept later evolved into what is now widely known as the "squeezed state."

A concrete mathematical formulation describing the effects of introducing such equivalence classes into a measurement system was derived in a 1977 paper [16] 
(see Figure 7). This technique is known as generalized heterodyne detection. 
However, it remained unknown whether the new degrees of freedom in the measurement process offered any advantages.

In the past, the invention of balanced homodyne detection by Yuen and Chan [71] has facilitated 
active discussion regarding methods for applying these concepts to practical optical signal processing systems [72]. 
In this study, we demonstrated that generalized heterodyne techniques can offer new capabilities for the application of squeezed states.

\section*{\textbf{C: Mechanism of auto control system with generalized heterodyne}}
The automatic control system adopted here is analogous to techniques of differential phase shift keying used in classical communication theory to mitigate 
the effects of channel instability—specifically, instability treated here as a signal phase shift. Because the channel's perturbing effects (phase shift as signal) 
act on both modes simultaneously, the relationship between the ellipse axes (or the phase difference) of the two modes remains invariant. 
By employing realization theorem—such as that found in Theorem 2—the projection effect of a generalized heterodyne POVM to signal state 
can be reproduced as squeezed states of two independent modes. One mode can then serve as the signal, while the other acts as the auxiliary quantum state 
for the generalized heterodyne mechanism. 

The combination of analogue of DPSK and the theorem 2 renders the relationship between the ellipse axes of the two modes unbiased, creating a system 
where the squared inner product of the target signal squeezed state and the generalized heterodyne POVM is determined by that of the initial state. 
We have demonstrated that this approach is effective for achieving the Monras-Olivares-Paris bound.

\section*{\textbf{D: Discussion on advantage of auto control}}
In this appendix, we discuss the advantages of the method proposed in this paper over the engineering approaches previously explored for achieving 
the Monras-Olivares-Paris bound.

[1] In estimation theory, the characteristics of quantum noise—which determine the ultimate 
quantum bound—are projected from the quantum space into a classical signal space via quantum measurement; this projection is described by 
an affine functional. The initial step in quantum estimation theory based on the separation theory involves applying concepts from classical statistics to 
this affine functional. It is a framework represented as ``Type B" in Figure 2 of this paper. In this approach, once a quantum estimator is determined, 
the corresponding quantum measurement is fixed, and the subsequent estimation process follows the same logic as classical statistical theory. 
We shall refer to this as the ``A-method."
Within this framework, addressing the challenge of ``efficient estimation"—in the context of formulating specific estimation problems via the Cramér-Rao
 inequality—presupposes the availability of a large number of copies of the quantum probe state. This is because achieving efficient estimation by A-method
  generally requires repeated measurements. The limiting case of this is asymptotic estimation theory, which necessitates the framework of
   ``quantum LAN" (Local Asymptotic Normality) theory, where an infinite number of copies are required.

[2] Next, we consider the Monras-Olivares-Paris bound, which is the focus of this paper. Discovered within the fields of quantum physics and mathematics, 
this bound is highly compelling, prompting many researchers to explore methodologies for achieving it—an endeavor that essentially becomes an engineering
 challenge. However, existing approaches have thus far relied on the A-method described in [1]. Consequently, the number of copies of the quantum probe state
  required to achieve efficient estimation is equal to the number of measurements needed for the efficient estimation. 
  This results in a significant delay in processing time.

[3] Here, the limitations of the aforementioned research become clear. From an engineering perspective, latency is a significant issue. 
Therefore, it is crucial to investigate methods for achieving the theoretical bounds—such as those predicted for "one-shot" measurements—
without incurring such latency.

[4] The method proposed in this paper is fundamentally based on the approach mentioned in [1], with one key distinction: it introduces a mechanism to control
 an affine functional, independent of the theory used to determine the quantum estimator. In other words, it incorporates a novel technique for controlling 
 quantum noise—a point that clarifies the title of this paper. We shall refer to this as the ``B-method."
In the case of the Monras-Olivares-Paris bound—used here as an example—even if the B-method is adopted, it was impossible to reach this bound 
when the control scheme is independent of the signal. This fact was explained in Section 6-1. To overcome this drawback, we introduce an automatic control
 technique so called here modified B-method—drawing an analogy to communication technology like differential phase modulation or Manchester encoding 
 (see Appendix C). This approach requires only two quantum state resources: the signal state and an auxiliary quantum state used for 
 generalized heterodyne detection. Consequently, the Monras-Olivares-Paris bound can be nearly achieved through a single shot measurement. 
 Thus, we do not need any other resource such as many copies of signal state.

The specifications for the required auxiliary quantum state were discussed in Section 6-2. 
 That is, the auxiliary state should be a squeezed state with the same energy as the signal squeezed state, 
which is then fed in parallel into the same modulation device (where phase rotation is induced by an external force). It is a vacuum squeezed state, and 
the average photon number is $ |\sinh(2z)|^2$, where $2z$ is a squeezing parameter.

\end{document}